\documentclass[journal=jctcce,manuscript=article]{achemso}
\setkeys{acs}{articletitle=true}

\usepackage[version=3]{mhchem} 
\usepackage{color}
\usepackage[bottom]{footmisc}

\def\H2{H$^{+}_{2}$}
\DeclareMathOperator{\e}{e}
\def\refe#1{{\color{blue}\text{Eq.}\,(\ref{#1})}}
\def\reff#1{{\color{blue}\text{}\,(\ref{#1})}}

\author{Marie Labeye}

\affiliation[]{Laboratoire de Chimie Physique Mati\`ere et Rayonnement, Sorbonne Universit\'e and CNRS, F-75005 Paris, France}
\altaffiliation{Contributed equally to this work}

\author{Felipe Zapata}
\affiliation[]{Laboratoire de Chimie Th\'eorique, Sorbonne Universit\'e and CNRS, F-75005 Paris, France}
\altaffiliation{Contributed equally to this work}

\author{Emanuele Coccia}

\affiliation{Dipartimento di Scienze Chimiche, Universit{\`a} di Padova,  Padova, Italy}

\author{Val\'erie V\'eniard}

\affiliation{ Laboratoire des Solides Irradi{\'e}s, \'Ecole Polytechnique, Universit\'e Paris-Saclay, CEA-DSM-IRAMIS, F-91128 Palaiseau, France}

\author{Julien Toulouse}

\affiliation[]{Laboratoire de Chimie Th\'eorique, Sorbonne Universit\'e and CNRS, F-75005 Paris, France}

\author{J\'er\'emie Caillat}

\affiliation[]{Laboratoire de Chimie Physique Mati\`ere et Rayonnement, Sorbonne Universit\'e and CNRS, F-75005 Paris, France}

\author{Richard Ta\"ieb}

\affiliation[]{Laboratoire de Chimie Physique Mati\`ere et Rayonnement, Sorbonne Universit\'e and CNRS, F-75005 Paris, France}

\author{Eleonora Luppi}

\affiliation[]{Laboratoire de Chimie Th\'eorique, Sorbonne Universit\'e and CNRS, F-75005 Paris, France}

\email{eleonora.luppi@upmc.fr}

\title{On the optimal basis set for electron dynamics in strong laser fields: \\ The case of molecular ion H$^{+}_2$}

\begin{document}

\begin{abstract} 

A clear understanding of the mechanisms that control the electron dynamics in strong laser field is still a challenge that requires to be interpreted by advanced theory. Development of accurate theoretical and computational methods, able to provide a precise treatment of the fundamental processes generated in the strong field regime, is therefore crucial. A central aspect is the choice of the basis for the wave-function expansion. Accuracy in describing multiphoton processes is strictly related to the intrinsic properties of the basis, such as numerical convergence, computational cost, and representation of the continuum. By explicitly solving the 1D and 3D time-dependent Schr\"odinger equation for H$^{+}_{2}$ in presence of an intense electric field, we explore the numerical performance of using a real-space grid, a B-spline basis, and a Gaussian basis (improved by optimal Gaussian functions for the continuum). We analyze the performance of the three bases for high-harmonic generation and above-threshold ionization for H$^{+}_{2}$. In particular, for high-harmonic generation, the capability of the basis to reproduce the two-center interference and the hyper-Raman phenomena is investigated. 
\end{abstract}


\section{Introduction}
\noindent

The optical response of a molecular system to an intense and ultrashort laser pulse is a subject of increasing interest since the advent of the attosecond laser pulses \cite{Chininatph2017}. Recent advances in laser technology are continuously triggering the introduction of new time-resolved spectroscopies, offering the opportunity to investigate electron dynamics in molecules with unprecedented time resolution\cite{KraStonatp2014}. For example, electronic charge migrations have been traced in molecules using attosecond pulses \cite{Lepine+nphot14}, electron correlation effects have been also observed in photoemission processes on the attosecond scale \cite{ossinatp2017,berguesnatcom2012} and above-threshold ionization (ATI) together with high-harmonic generation (HHG) spectra have been used to explain the attosecond dynamics of electronic wave packets in molecules \cite{nisochemrev2017,Haessler:2010hb}. 

Despite these exciting experimental achievements, reaching a clear understanding of the mechanisms that control the electron dynamics under the action of a strong laser field is still a challenge that requires theoretical support\cite{nisochemrev2017}. It is crucial to develop accurate theoretical and computational methods capable to provide precise treatments of the fundamental processes generated by a strong laser field\cite{PalaSanVMartTopRev15,PhysRevA.76.043412,PhysRevA.77.013414,PhysRevA.76.043419}.

Nowadays, the electron dynamics problem in strong fields is tackled by two main families of methods: time-dependent density-functional theory (TDDFT) and time-dependent wave-function
methods\cite{nisochemrev2017,coccia16b,Dinh2017,PhysRevA.94.033410,IngaNest2011,Chu2005jcp}. With these methods, developments have been focused on the accurate description of electron correlation. However, because of the complexity of nonlinear optical phenomena, such as HHG and ATI, another important aspect needs to be carefully addressed: the choice of the one-electron basis  for representing the time-dependent wave function. In fact, a reliable description of the electron dynamics in strong laser fields depends on the accuracy in reproducing the bound states and, even more important, the continuum states of the molecular system considered. In addition, choosing a good basis can improve the numerical convergence of the results and reduce the computational cost of simulations.

Most of the proposed numerical methods in literature directly describe the system wave function on a real-space grid\cite{krause_calculation_1992,Wassaf+pra03,Ruiz:2006bz,PhysRevA.93.023434} or through a numerically defined grid-based basis set of functions, as in the case of the discrete-variable representation method\cite{tao09}, the pseudospectral grid method, or the finite-element method\cite{PhysRevA.94.033421}. Within these approaches, schemes have been proposed to compute ATI spectra in molecules\cite{PhysRevA.85.062515} and to study the different molecular orbital contributions to HHG spectra\cite{PhysRevA.93.013422,Wang:17}. Grid-based basis sets have demonstrated to be very accurate to describe nonlinear optical phenomena. However, the computational cost can be very high and strategies involving multi-level parallelization schemes have had to be developed\cite{Octopus2015}. 

Another recurrent basis, in the context of ultrafast electron dynamics, is composed by B-splines, defined as piecewise polynomial functions with compact support\cite{Boor:78}. They were first introduced in atomic calculations by Shore\cite{Shore:73} and later extensively used to treat ionized and excited states\cite{CFFischer89,CFFischer90}. B-splines have proved to be a very powerful tool to describe multiphoton ionization processes in atoms and molecules in the frameworks of TDDFT and wave-function methods\cite{fernandoJPB1999,bachaurep2001,ECormier97,Stener07}. The success of B-splines is due to a remarkable feature: B-splines are able to reproduce accurately both bound and continuum states. This numerical property is directly related to their effective completeness\cite{Argenti09}. Nowdays atomic packages based on B-splines are available\cite{CFFischer11,Nikolopoulos03,MForre10} and recent studies show their ability to reproduce HHG and ATI spectra of molecules under the action of a strong laser field\cite{BFetic17}. However, new algorithms have to be developed in order to increase the computational efficiency of complex calculations with B-splines.

More recently, Gaussian-type orbital functions (abbreviated as Gaussian functions in the following), in the framework of the time-dependent configuration-interaction (TDCI) method, have been used to calculate HHG spectra in atoms and molecules\cite{lupp+13jcp,white15,coccia16b,coccia16a,lupp+12mol}. The importance of the cardinal number (related to the maximal angular momentum) of the basis set and the number of diffuse basis functions was investigated.\cite{lupp+13jcp,coccia16b} Two strategies to improve continuum states have been studied: multi-centered basis functions\cite{white15,coccia16b} and, alternatively, Gaussian functions with exponents specially optimized to improve the continuum\cite{coccia16a,CocAssLupTou-JCP-17}. This latter strategy proved to be more efficient than using multi-centered basis functions and it has also lower computational cost, however it remains to be tested on molecular systems. These works permitted us to identify the best basis sets to be used in order to capture the features of HHG spectra.

Finally, to overcome some of the limitations of the grid, B-spline, and Gaussian basis, hybrid approaches have been proposed in the last years. For example, Gaussian functions were used together with grid-based functions to reproduce electron dynamics in molecular systems\cite{Yip2014pra}, and also Gaussian functions have been combined with B-splines for studying ionization in H and He atoms\cite{mara14pra,marantejctc2017}.
 
The aim of the present work is to compare the performance of the three families of basis, briefly reviewed above, i.e. grid, B-splines, and Gaussians, for the calculation of HHG and ATI spectra of the molecular ion H$^{+}_{2}$. This system has been chosen because it has the advantage of having only one electron, which allows us not to bias our investigation with possible effects due to electron correlation. Indeed, with this simple case, we can focus on the effectiveness of the representation of the continuum states for the electron dynamics and the computational advantages of each basis. Moreover, the presence of two nuclei in H$^{+}_{2}$ offers the opportunity to observe intricate physical features, such as quantum interferences in the HHG process\cite{wor10,APicon11,PhysRevA.66.023805}.

This article is organized as follows. In Section \ref{theory1D} we present the 1D theoretical model to solve the electronic time-dependent Schr\"odinger equation (TDSE) with grid, B-spline, and Gaussian bases. In Section \ref{results1D} we present and discuss the results for the 1D approach. In Section \ref{theory3D} we present the 3D theoretical model to solve the electronic TDSE with grid and Gaussian basis. In Section \ref{results3D} we present and discuss the results for the 3D approach. We compare the bound and the continuum energy spectra of H$^{+}_{2}$, as well as HHG and ATI spectra for grid, B-spline, and Gaussian bases, emphasizing the advantages and disadvantages of each representation. In particular, for HHG spectra, we investigate the capability of the different basis to reproduce specific quantum features, such as the hyper-Raman \cite{mil93} and the the two-center interference phenomena\cite{wor10,APicon11,PhysRevA.66.023805}. Finally, Section \ref{conclusions} contains our conclusions.

\section{1D theoretical model of H$^{+}_{2}$}
\label{theory1D}
\noindent

The electronic TDSE for a 1D model of H$^{+}_{2}$ is given by, in atomic units (au),
\begin{equation}
\label{TDSE}
i\frac{\partial}{\partial t}\psi(x,t) = \left[\hat{H}_{0}(x) + \hat{H}_\mathrm{int}(x,t) \right] \psi(x,t),
\end{equation}
where $\psi(x,t)$ is the time-dependent electron wave function. Here, $\hat{H}_0(x)$ is the field-free Hamiltonian,
\begin{equation}
\label{free}
\hat{H}_0(x)= -\frac{1}{2} \frac{\mathrm{d}^{2}}{\mathrm{d}x^{2}}+\hat{V}(x),
\end{equation}
with a soft Coulomb electron-nuclei interaction given by
\begin{equation}
\label{soft}
\hat{V}(x)=-\frac{1}{\sqrt{\left(x-\frac{R}{2}\right)^2+\alpha}} -\frac{1}{\sqrt{\left(x+\frac{R}{2}\right)^2+\alpha}},
\end{equation}
where $R$ is the interatomic distance and $\alpha$ is a parameter chosen to reproduce the exact ionization energy $I_\mathrm{p}$ (taken as -1.11 Ha for all the three bases employed here) of the real H$^{+}_{2}$ molecule at a given value of $R$ ($\alpha=1.44$ at $R$ = 2.0 au)\cite{PhysRevA.66.023805}.

The interaction between the electron and the laser electric field $E(t)$ is taken into account by the time-dependent interaction potential, which is given in the length gauge by
\begin{equation}
\hat{H}_\mathrm{int}(x,t)=\hat{x}E(t),
\end{equation}
where $E(t)$ is the laser electric field and $\hat{x}$ is the electron position operator. The laser electric field is chosen as $E(t)=E_0 f(t) \sin(\omega_0 t)$ where $E_0$ is the maximum amplitude of the pulse, $\omega_0$ is the carrier frequency, and $f(t)$ is a trapezoidal envelope
\begin{equation}
f(t) = \Biggl\{
 \begin{array}{lll}
  t/T_0                  ,&\;\;0&\leq t<T_0\\
  1                      ,&\;\;T_0&\leq t<9T_0 \\
  10-t/T_0               ,&\;\;9T_0&\leq t < 10T_0,
 \end{array}
\end{equation}
with $T_0=2\pi/\omega_0$. The duration of the pulse is thus $\tau = 10 T_0$ (i.e., 10 optical cycles).

\subsection{HHG and ATI spectra}

A HHG spectrum, experimentally accessible by measuring the emission spectrum in the presence of an intense laser field, can be calculated as the acceleration power spectrum over the duration of the laser pulse $\tau$\cite{bur92}
\begin{equation}
\label{HHGdipole}
P_\text{a}(\omega) =  \left| \int_0^{\tau} \left\langle \psi(t) \vert -\nabla \hat{V} - E(t) \vert \psi(t)\right\rangle W(t) e^{-i\omega t} \mathrm{d}t \right|^2,
\end{equation}
where $-\nabla \hat{V} - E(t)$ is the electron acceleration operator, as defined by the Ehrenfest theorem, and $W(t)$ is an apodisation function that we chose to be of the sine-square window form. An alternative way to obtain the HHG spectrum is to calculate the dipole power spectrum as
\begin{equation}
\label{HHGacc}
P_\text{x}(\omega) =  \left| \int_0^{\tau} \left\langle \psi(t) \vert \hat{x} \vert \psi(t)\right\rangle W(t) e^{-i\omega t} \mathrm{d}t \right|^2,
\end{equation}
It can be shown that the two forms are related \cite{bur92,ban09,Han10,coccia16b}, $\omega^{4} P_\text{x}(\omega) \approx P_\text{a}(\omega)$, under reasonable conditions (see Appendix in Ref. \cite{coccia16b}).  The function $W(t)$ is a sin-square window function chosen empirically to minimise the noise, and especially to remove the artefacts arising from the discrete Fourier transform due to the fact that we integrate only over a limited time duration and not from $-\infty$ to +$\infty$.

An ATI spectrum, which is experimentally accessible by measuring the photoelectron spectrum of the molecule, can be calculated by spectrally analyzing the system wave function $\psi(\tau)$ at the time $\tau$ corresponding to the end of the laser pulse. Specifically, using the window operator method, one calculates the probability $P(E,n,\gamma)$ to find the electron in the energy interval $[E-\gamma,E+\gamma]$ as  \cite{SCHAFER91,ati}
\begin{equation}
\label{PHOTO}
P(E,n,\gamma) = \left\langle \psi(\tau) \left| \frac{\gamma^{2^n}}{(\hat{H}_ 0-E)^{2^n} + \gamma^{2^n}} \right| \psi(\tau)\right\rangle,
\end{equation}
where $\gamma$ and $n$ are parameters chosen to allow flexibility in the resolution and accuracy of the energy analysis. In our case we chose $n=2$ and $\gamma=2\times10^{-3}$ au.

\subsection{Representation of the time-dependent wave function and propagation}

\subsubsection{Real-space grid} 

The time-dependent wave function is discretized on a real-space grid of $N$ points $x_i$ separated by a constant step $\Delta x$ = $x_{i+1} - x_{i}$, in the interval [$x_{1} = -(N-1)\Delta x/2 ,  x_{N} =  (N-1)\Delta x/2$]. It is thus represented by the vector
\begin{equation}
\psi(x,t) \equiv (\psi(x_1,t),\ldots,\psi(x_i,t),\ldots,\psi(x_N,t)),
\label{gridpsi}
\end{equation}
where $x_i = (i-1-(N-1)/2)\Delta x$. 

The Laplacian operator is computed with the second-order central difference formula which gives rise to a tridiagonal matrix representation of the Hamiltonian $\hat{H}_0$\cite{krause_calculation_1992}. The TDSE (\refe{TDSE}) is solved by means of the Crank-Nicholson propagation algorithm\cite{crank_nicolson}. The H$^+_2$ ground state, computed by inverse iteration\cite{numerical_recipies}, is taken as the initial state for the propagation. In addition, to avoid unphysical reflections at the boundaries of the simulation grid, a mask-type absorber function\cite{krause_calculation_1992} was implemented with a spatial extension of $50$ \text{au}. 

For ATI spectra, converged results were obtained with $N=200001$ and $\Delta x=0.02$ \text{au}, and with a time step $\Delta t = 8.41\times10^{-4} $ \text{au}. For HHG spectra, we obtained converged results with $N=160001$, $\Delta x=0.01$ \text{au}, and $\Delta t=1.35\times10^{-2}$ \text{au}.

\subsubsection{B-spline basis set}

The time-dependent wave function with the B-spline basis set is represented as 
\begin{equation}
\psi(x,t) = \sum_{i=1}^{M}c_i(t) B_{i}^{k}(x),
\end{equation}
where $c_i(t)$ are time-dependent coefficients and $\{B_{i}^{k}(x)\}$ are a set of B-spline functions of order $k$ and dimension $M$. To completely define B-spline functions a sequence of knots ${\bf t} = \{t_i\}_{i=1,M+k}$ must be given. Each function $B^{k}_i(x)$ is defined on a supporting interval $[t_i,t_{i+k}]$ which contains $k+1$ consecutive knots, and the function $B^{k}_i(x)$ vanishes outside this interval. We have chosen the first and the last knots to be $k$-fold degenerate, $t_1 = t_2 = \cdots = t_{k} = R_{\text{min}}$ and $t_{{M+1}} = t_{{M+2}} = \cdots = t_{{M+k}}= R_{\text{max}}$, while the multiplicity of the other knots is unity. The width of an interval is $t_{i+1} -  t_{i} = R_{\text{max}}/(M-k+1)$\cite{bachaurep2001}. In our calculations we used $k=8$, $M=15008$, $R_{\text{min}}=0$, and $R_{\text{max}} = 8000$ \text{au}. The system was placed at the center of the box at $x$ = 4000 \text{au}. 

ATI and HHG spectra were obtained by solving the TDSE (\refe{TDSE}) within the Cranck-Nicholson propagation algorithm\cite{crank_nicolson} using a time step of $\Delta t=1.35\times10^{-2}$ \text{au}. The H$^+_2$ ground state was computed by inverse iteration\cite{numerical_recipies} and taken as the initial state for the propagation. We did not need to use any absorber during the propagation because of the very large size of the simulation box. 

\subsubsection{Gaussian basis set}

For the Gaussian basis set we followed the TDCI procedure developed in our previous work\cite{coccia16b}, and adapted it to the present 1D H$^+_2$ model. The time-dependent wave function is represented here as 
\begin{equation}
\psi(x,t) = \sum_{k \geq 0} {c}_{k}(t) \phi_{k}(x),
\end{equation}
where $\phi_{k}(x)$ are the eigenstates of the field-free Hamiltonian $\hat{H}_0$, composed by the ground state ($k=0$) and all the excited states ($k>0$). The $\phi_{k}(x)$ are expanded on the Gaussian basis set. In this work, we use uncontracted Gaussians localized on each nucleus and two ``angular momenta'' ($\ell$), corresponding to odd and even functions. The basis functions are thus of the form $(x\pm R/2)^\ell\e^{-\alpha (x \pm R/2)^2}$, where $\ell=0$ or $1$. The Gaussian exponents $\alpha$ are of two different types. The first type of exponents are optimized to describe the bound part of the wave function. We used the {uncontracted} STO-3G basis set, i.e. three uncontracted Gaussians whose exponents are taken from the STO-3G basis set with Slater exponent $\zeta=1$. We take the same exponents for $\ell=0$ and $\ell=1$. The second type of exponents are optimized for the representation of the continuum\cite{coccia16b}. They are computed with the procedure developed by Kaufmann\cite{kauf+89physb} adapted to the 1D model, i.e. by optimizing the overlap between a 1D Slater type function $N^{(S)}_n(\zeta) x^n\e^{-\zeta |x|}$ with $\zeta=1$ and a Gaussian function $N^{(G)}_\ell(\alpha_{n,\ell}) x^\ell\e^{-\alpha_{n,\ell}x^2}$, where $N^{(S)}_n$ and $N^{(G)}_\ell$ are normalization factors. Note that, in this case, the exponents used for the $\ell=0$ shell and for the $\ell=1$ shell are different. In the following, we will denote these Gaussian functions optimized for the continuum as K functions. To sum up, we use 3 functions with STO-3G exponents and 4 K functions for each angular momentum, localized on each nucleus, which makes a total of $(3+4)\times4=28$ uncontracted Gaussian basis functions. However when we orthonormalize this basis set, we find linear dependencies that needs to be removed. For this we define a cutoff $\epsilon=10^{-8}$ under which the eigenvalues of the overlap matrix are considered to be zero, and their corresponding eigenvectors are removed from the space. We get an orthonormalized basis set of 24 basis functions. The basis-set exponents are collected in Table S1 of Supporting Information.
To solve the TDSE (\refe{TDSE}) we used the split-operator propagator with $\Delta t=1.35\times10^{-2}$ \text{au}. 

In order to compensate for the unphysical absence of ionization, we used the double-$d$ heuristic lifetime model proposed in\cite{coccia16b}. This model requires two parameters: $d_0$ and $d_1$ which represent different electron escape lengths after ionization. We have chosen these parameters on the basis of the rescattering model\cite{cork93prl,lewe+pra94} where an electron is ionized by a strong laser field, accelerated in the continuum, and then brought back close to its parent ion where it can recombine or scatter. From this model, $d_0$ is equal to the maximum electron excursion after ionization which is $x_{\text{max}}=\sqrt{{2E_0}/{\omega_0^4}}$, while $d_1 < d_0$. In our calculations we always used $d_1$ = 20 au. Moreover $d_0$ affects all the continuum states below the cutoff energy $E_{\text{cutoff}}=I_\text{p}+3.17U_\text{p}$\cite{cork93prl,lewe+pra94} ($U_\text{p}=E_0^2/(4\omega_0^2)$ is the ponderomotive energy of the electron) while $d_1$ handles the ionization for those continuum energy states above $E_\text{cutoff}$. This allows to better retain the contribution of continuum states for the recombination step of the HHG process. \ref{tab1} collects the values of $d_{0}$ used in this work.\\

\begin{table}
\caption{$d_{0}$ values, taken as $x_\text{max}$, used in the double-$d$ heuristic lifetime model for the laser intensities employed in this work.
\label{tab1}}
\begin{tabular}{cc }
\hline \hline 
$I$ (W/cm$^{2}$) & $d_{0}$ (au) \\
\hline  
$5\times10^{13}$   & 23    \\
     $10^{14}$  & 33    \\ 
 $2\times10^{14}$  & 46   \\
 $3\times10^{14}$ & 57   \\
 $4\times10^{14}$ & 66 \\
 $5\times10^{14}$ & 74   \\
 $7\times10^{14}$ & 87    \\
\hline \hline 
\end{tabular}
\end{table}

There is a fundamental difference between this approach and the grid and B-spline ones. Indeed, the TDSE with the Gaussian basis set is solved in the energy space. This fact permits to have a more direct and intuitive interpretation of the role of bound and continuum states in HHG and ATI spectroscopies. In addition, the use of Gaussians reduces considerably the computational time required in time propagation. This makes it a more promising tool for the modelisation of larger molecules.

\begin{figure}[ht!]
\centering
\includegraphics[scale=0.7]{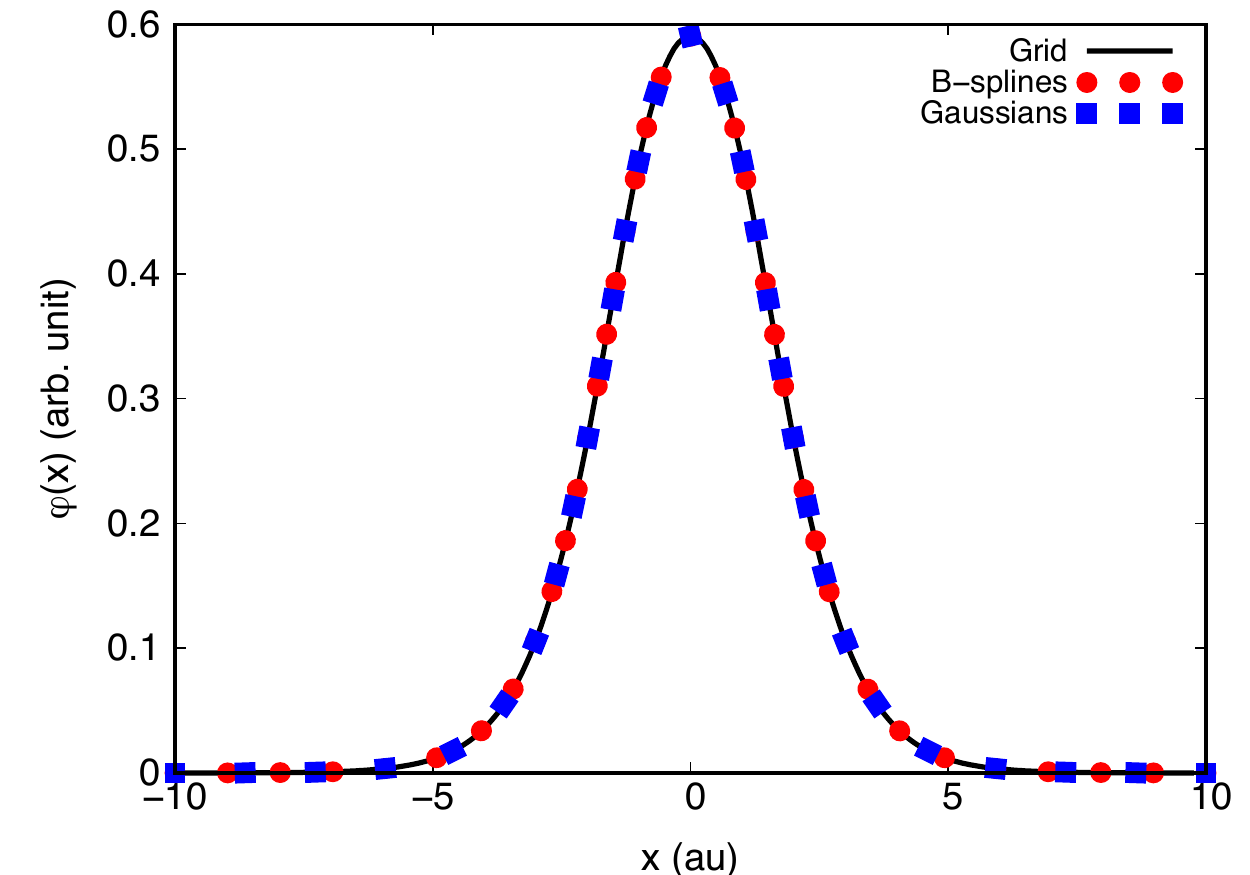}
\caption{Ground-state wave function of H$_2^+$ (at the equilibrium internuclear distance of $R=2.0$ au)} calculated using grid, B-spline, and Gaussian basis.
\label{ground}
\end{figure}

\section{1D RESULTS AND DISCUSSION \label{results1D}}

\subsection{Spectrum of the field-free Hamiltonian\label{field-freeH}}

The spectrum of $\hat{H}_0$ should be strictly independent on the choice of the basis set in the limit of a complete basis set. However, because our basis sets are not complete, differences in the eigenstates and eigenvalues from grid, B-spline, and Gaussian basis sets can arise, especially at high-energy values.  In order to investigate the behavior of the three basis sets,  the spectrum of $\hat{H}_{0}$ is analyzed in this section.

\begin{figure}[t]
\centering
\includegraphics[scale=0.7]{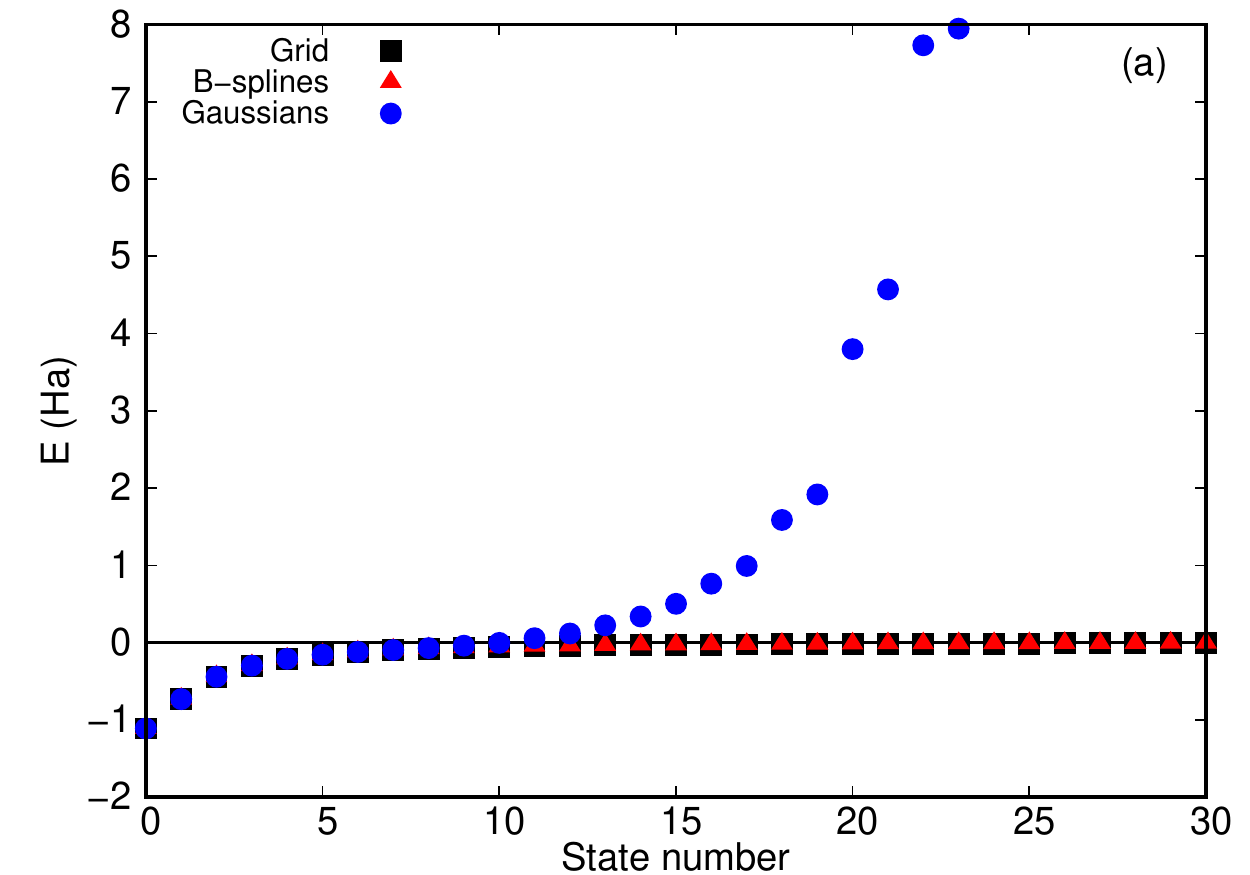}
\includegraphics[scale=0.7]{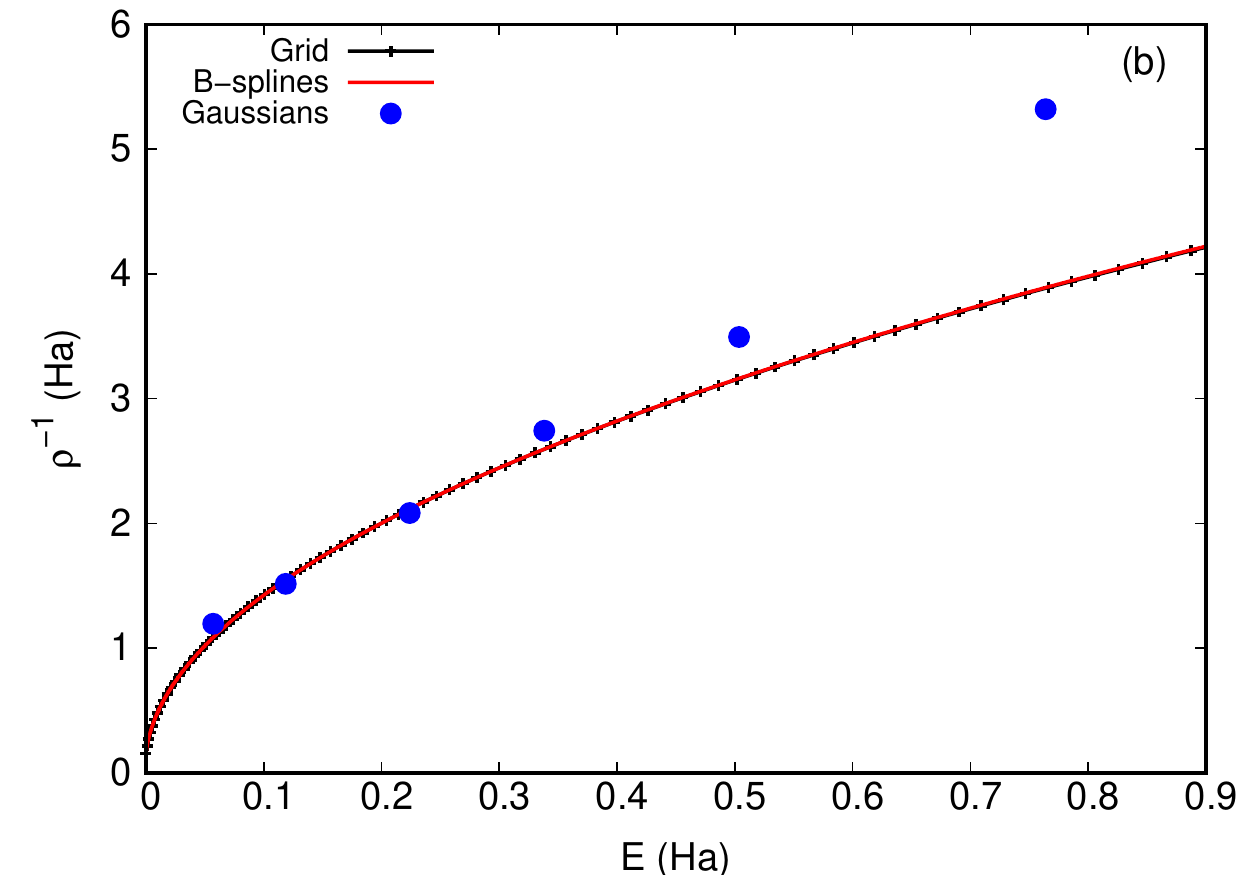}
\caption{(a) Eigenvalues of H$_{2}^{+}$ up to the 30th eigenstate. (b) Inverse of the normalized density of continuum states.}
\label{DOS}
\end{figure}

In \reff{ground} the ground-state wave function is shown. The three basis sets reproduce exactly alike the ground state of the 1D H$_2^+$ model, at the equilibrium internuclear distance of $R=2.0$ au. The panel (a) of \reff{DOS} shows the eigenvalues given by each basis set up to the 30th energy state, and in panel (b) of \reff{DOS} one finds the inverse of the density of continuum states which is defined as $\rho(E_j) = {1}/{(E_{j+1}-E_j)}$ where $E_j$ is a positive eigenvalue. In order to compare the three bases, the density of the states has been normalized to the length of the simulation box in the case of the grid and B-splines and to a constant in the case of the Gaussians. This constant was chosen to force the first Gaussian continuum eigenvalue to match the first continuum eigenvalue of the grid and B-splines, which are identical. For all the three basis sets, the continuum part of the spectrum is represented as a finite number of eigenstates as, in numerical calculations, the basis set is always incomplete. However, the discreteness of the Gaussians is much larger than that of the grid and B-splines. The spectrum obtained with the Gaussians starts to diverge from the grid and B-spline ones already at around the 13th state. This issue is a direct consequence of the relatively small size of the Gaussian basis set compared to the number of grid points or B-spline functions used. Indeed, the STO-3G+4K basis contains only 24 Gaussian basis functions whereas we used 400001 grid points and 15000 B-splines. In principle, we could increase the number of Gaussians but this will quickly lead to the linear dependency problem. This problem prevents us to use more than a few tens of optimized Gaussian functions. This fact, as we will see in the following sections, can have important consequences on the calculation of HHG and, in particular, of ATI spectra.

\begin{figure}[h!]
\centering
\includegraphics[scale=0.7]{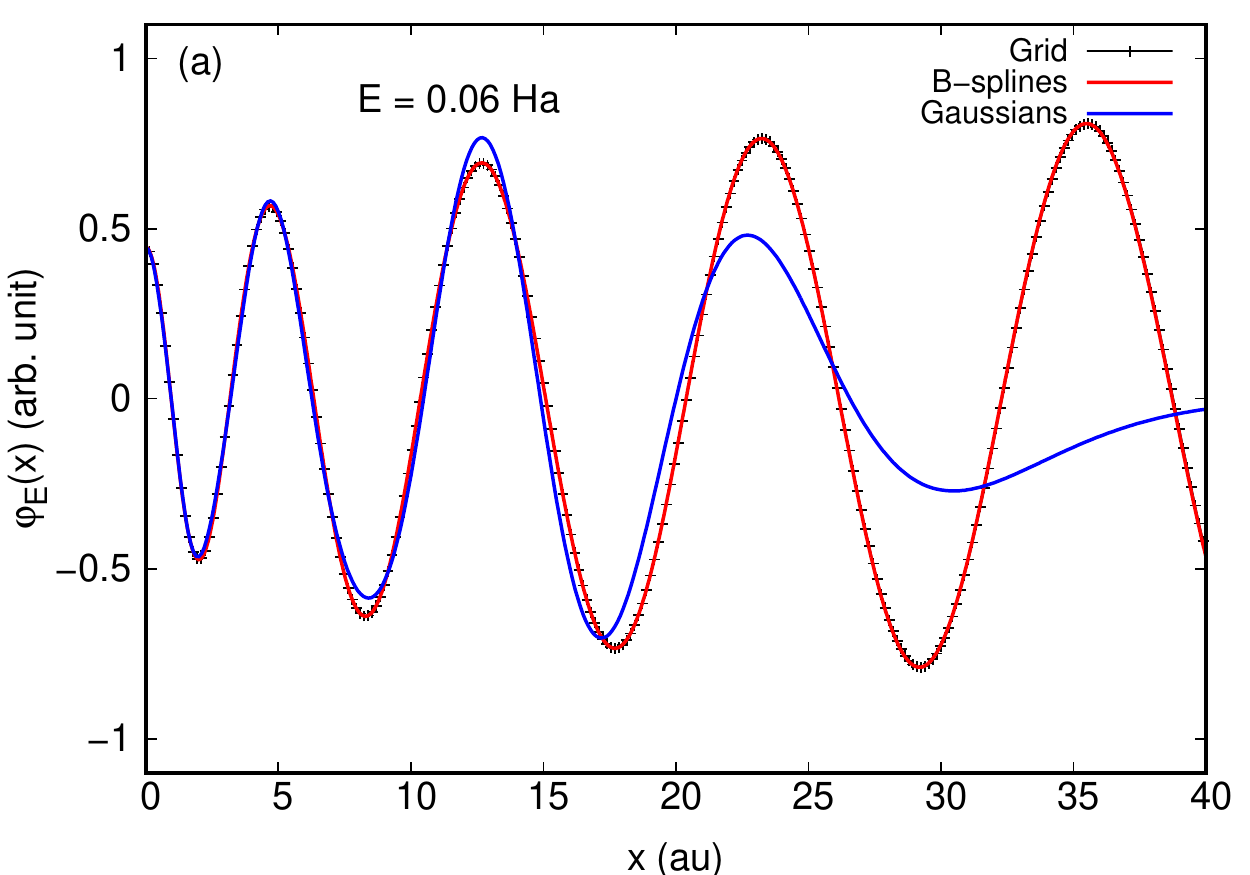}
\includegraphics[scale=0.7]{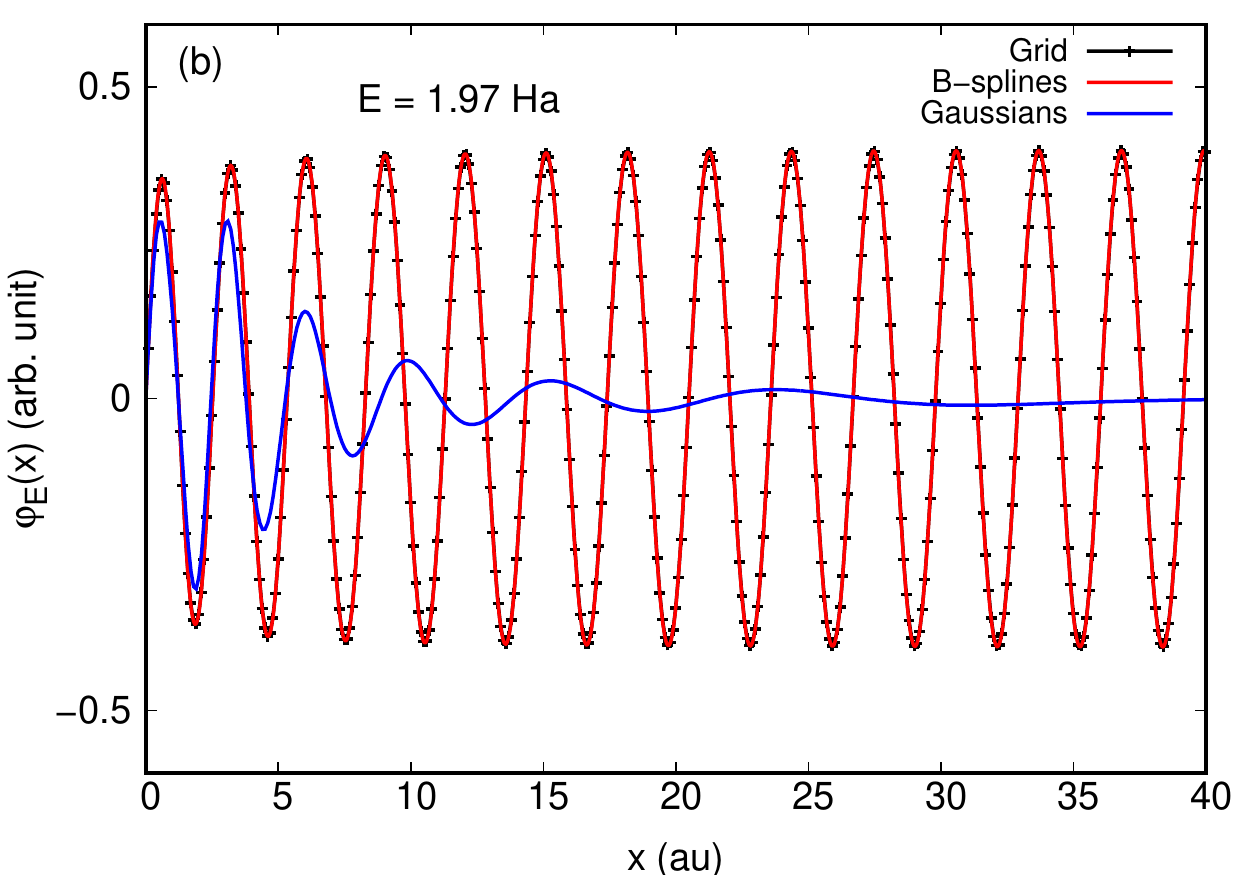}
\caption{(a) Spatial dependence of the even wave function $\varphi_E(x)$ corresponding to $E=0.06$ Ha. (b) Spatial dependence of the odd wave function $\varphi_E(x)$ corresponding to $E=1.97$ Ha.}
 \label{continuum}
\end{figure}

To investigate the accuracy of the grid, B-spline, and Gaussian bases in the description of continuum wave functions, we have chosen two different continuum energies, both representative of two different continuum energy regions: low energy ($E = 0.06$ Ha) and high energy ($E = 1.97$ Ha). For each of these energies, we reported in \reff{continuum} the corresponding wave functions $\varphi_{E}(x)$. For the grid, the continuum wave functions were obtained by propagating the TDSE at the chosen positive energy $E$ with a fourth-order Runge-Kutta algorithm\cite{numerical_recipies}, and then normalized with the Str\"omgren procedure\cite{stromgren}\cite{str}. Instead, for B-splines and Gaussians, the wave functions were obtained from a direct diagonalisation of $\hat{H}_0$. In this case, the resulting continuum states were renormalized using the procedure proposed by Mac\'ias \emph{et al.} \cite{Macias88}\cite{mac}. We verified that the Str\"omgren and Mac\'ias procedures are equivalent \cite{phd}. The continuum wave functions computed with both grid and B-spline basis sets reproduce the same oscillations in the low- and high-energy regions of the continuum. On the other hand, Gaussians can reproduce just a few of the oscillations. We already observed this behavior in the case of the hydrogen atom in a 3D calculation\cite{coccia16a} where the crucial role of the K functions was pointed out in order to obtain these oscillations (in that case a much larger basis set was employed). Here, we want to draw the attention on the fact that Gaussians can still be reasonable in the low-energy continuum, but become unsuitable to reproduce oscillations for high-energy continuum states. The probability of propagating an electron in one of the two regions depends on the laser parameters used in the simulation. This fact can have important implications in the description of HHG and ATI spectra as we will see in the following sections.

\subsection{HHG}

\begin{figure}
\centering
\includegraphics[scale=0.7]{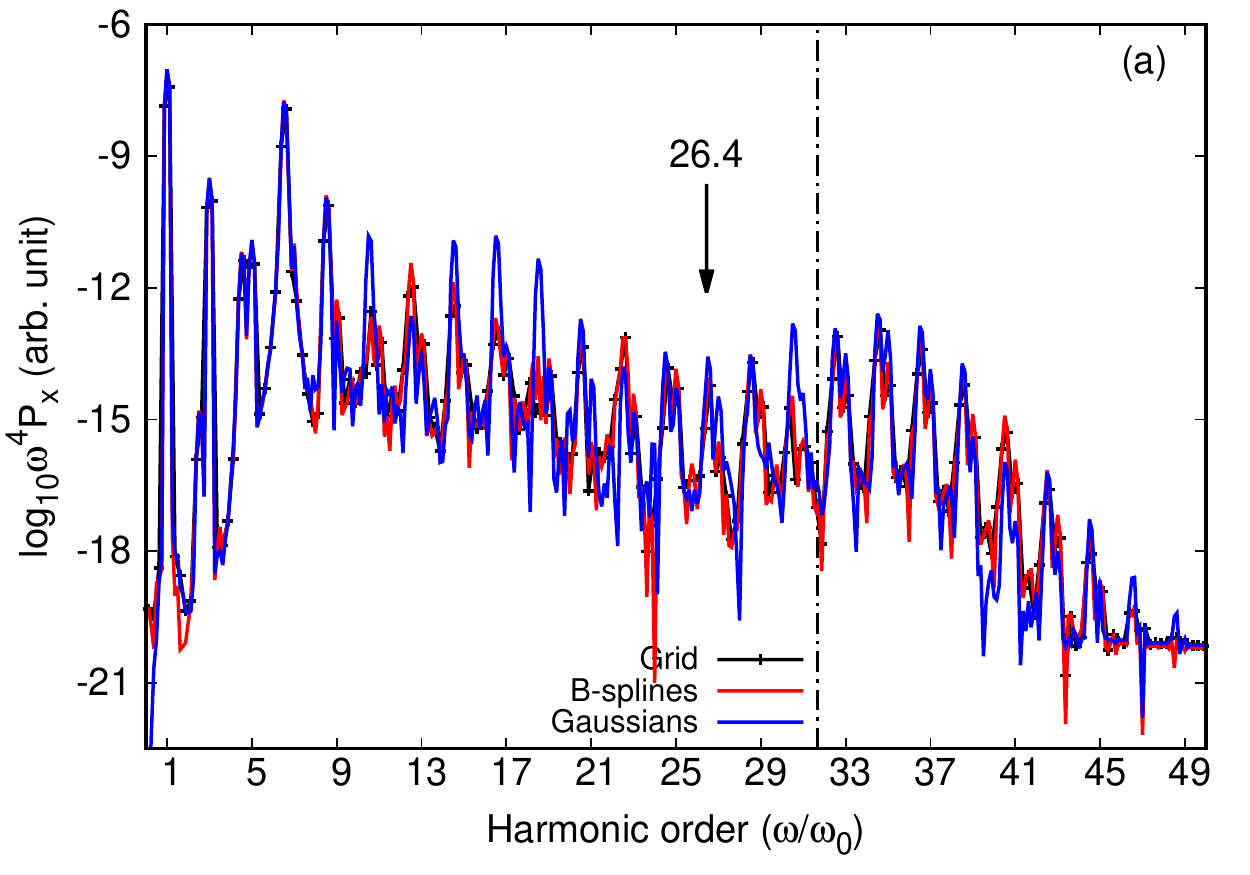} 
\includegraphics[scale=0.7]{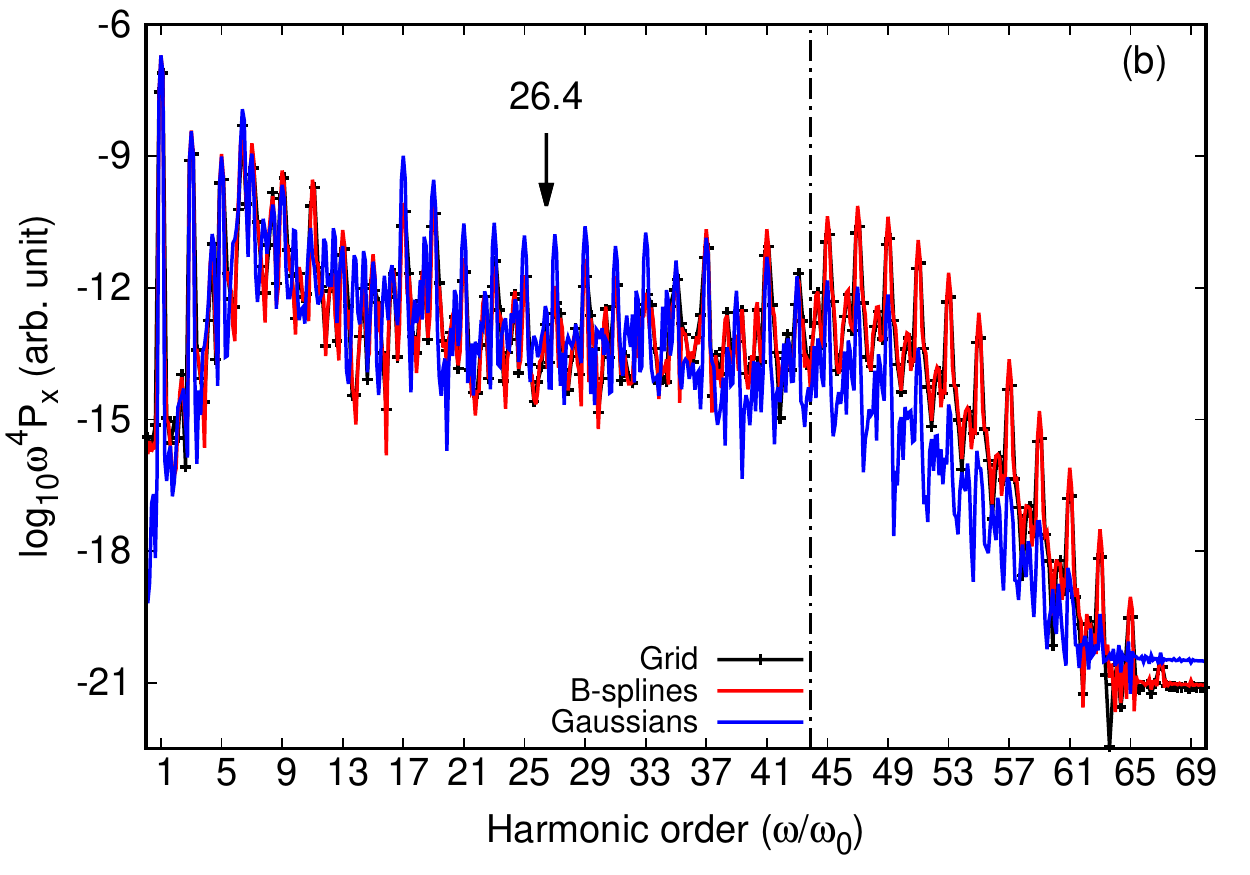} 
\includegraphics[scale=0.7]{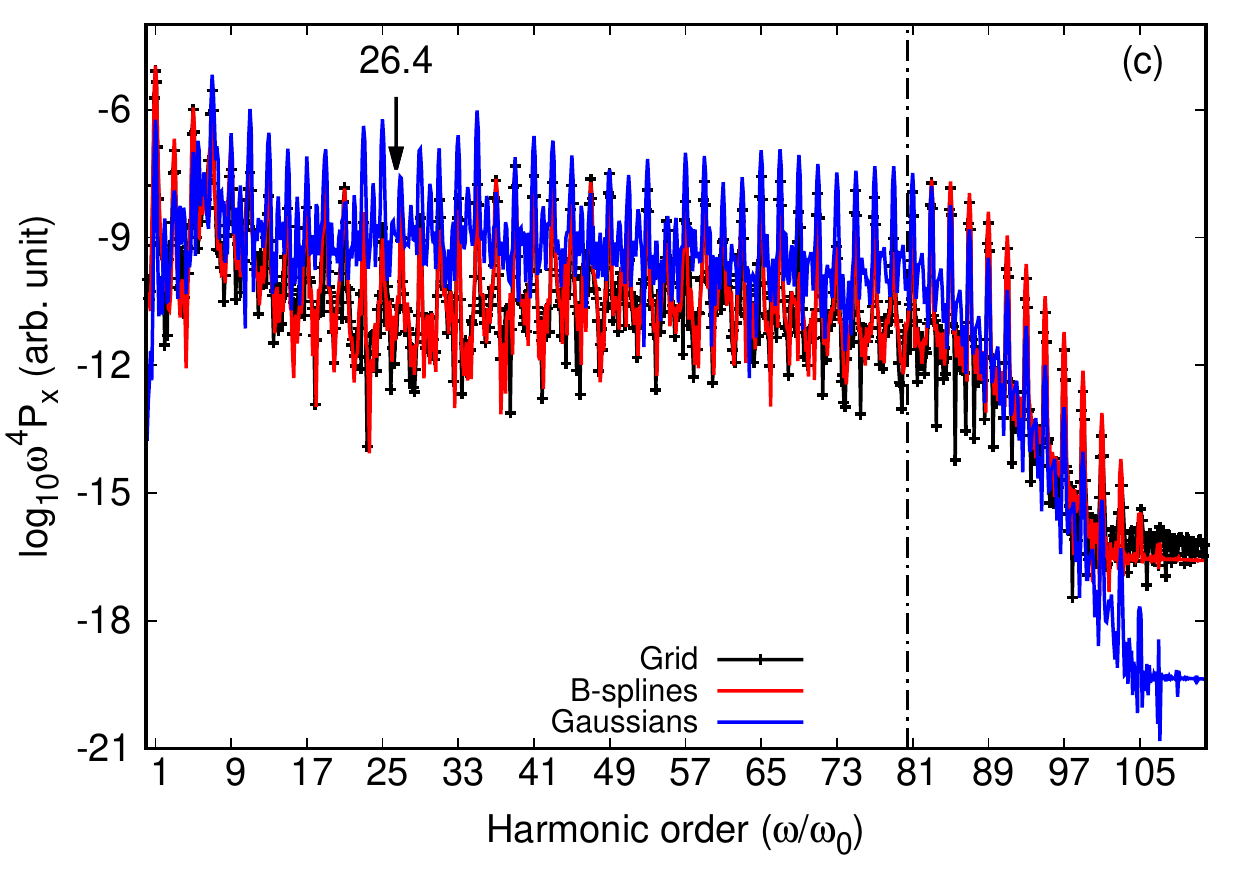} 
\caption{HHG spectra calculated from the electron dipole at the equilibrium internuclear distance $R$ = 2.0 au with laser intensities: (a) $I=10^{14}$~W/cm$^2$, (b) $I=2\times 10^{14}$~W/cm$^2$, and (c) $I=5 \times 10^{14}$~W/cm$^2$. 
Intensities $I=5\times 10^{13}$ and $7 \times 10^{14}$ W/cm$^2$ are reported in the Supplementary Information.  For each HHG spectrum, the dot-dashed lines indicate the cutoff energies, which are given by the rescattering model as $E_{\text{cutoff}}=I_\text{p}+3.17U_\text{p}$, see Ref. \cite{cork93prl,lewe+pra94}: (a) $E_{\text{cutoff}} = 31.7\omega_0$, (b) $E_{\text{cutoff}} = 43.9\omega_0$, and (c) $E_{\text{cutoff}} = 80.5\omega_0$. The arrow points to the expected position of the two-center interference minimum extracted from the recombination dipole.}
\label{hhgdip}
\end{figure}

\begin{figure}[h!]
\centering
\includegraphics[scale=0.7]{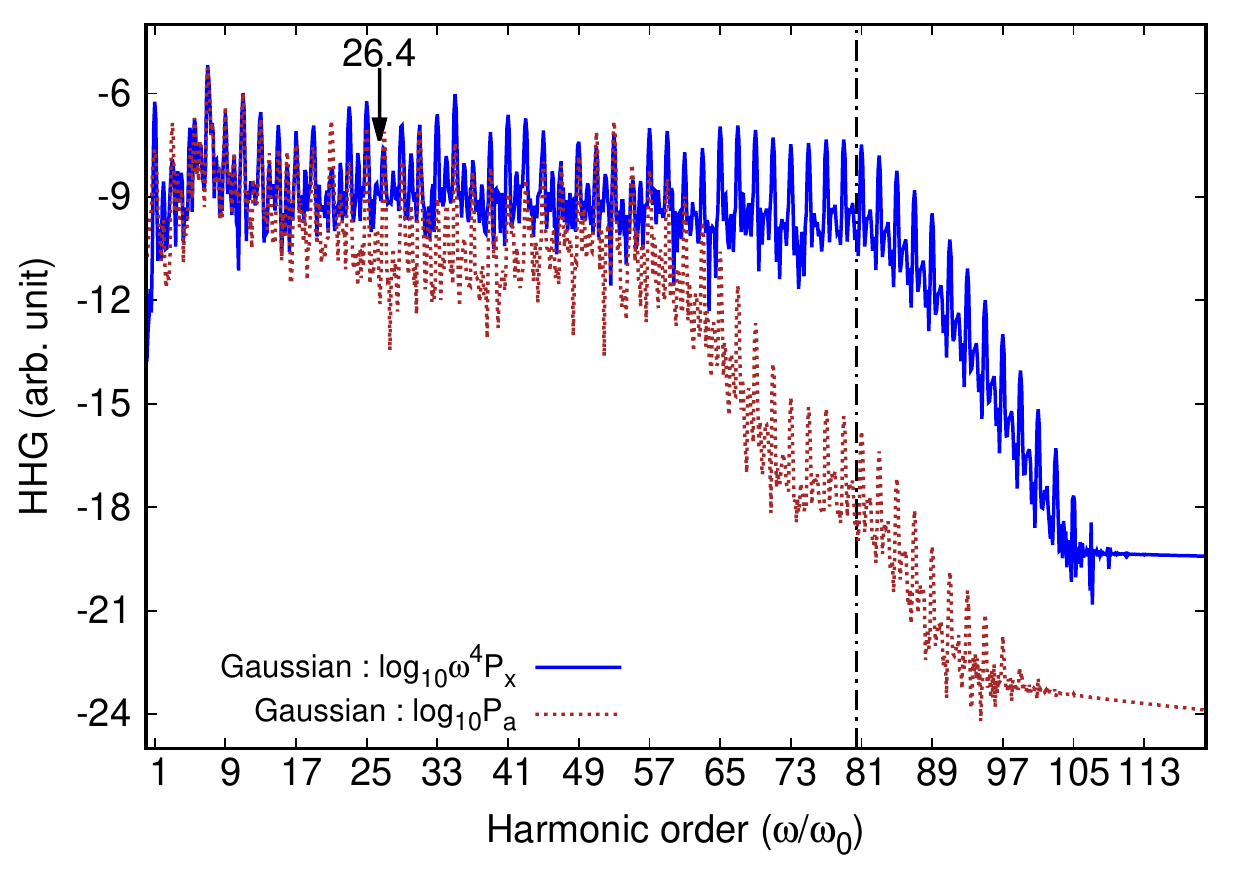} 
\caption{HHG spectra calculated from the electron dipole and the electron acceleration at the equilibrium internuclear distance of $R=2.0$ au with a laser intensity of $I=5 \times 10^{14}$ W/cm$^2$ using Gaussian basis sets. The dot-dashed line is the cutoff energy $E_{\text{cutoff}} = 80.5\omega_0$ and the arrow points to the expected position of the two-center interference minimum, extracted from the recombination dipole which is identical to the one extracted from the recombination acceleration.}
\label{hhgdipacc5d14}
\end{figure}

HHG spectra have been calculated in the dipole and the acceleration forms for H$_{2}^{+}$ at different internuclear distances: $R = 1.8$, $R=2.0$ (equilibrium distance), and $R=2.2$ au for a Ti:Sapphire laser pulse with a carrier frequency $\omega_0=0.057$ Ha (1.55 eV, 800 nm) and different intensities: $I=5\times10^{13}$, $I=1\times10^{14}$, $I=2\times10^{14}$, $I=5\times10^{14}$, and $I=7\times 10^{14}$ W/cm$^2$. 

In \reff{hhgdip} we show the dipole form of the HHG spectra at $R$ = 2.0 au for three different laser intensities. All the three basis sets reproduce the general expected features of an HHG spectrum: the intensity of the low-order harmonics decreases rapidly, then a plateau region follows where the intensity remains nearly constant, and at high frequencies the harmonic intensity decreases again. As H$_{2}^{+}$ has a center-of-inversion symmetry, only odd harmonics are presented in the spectrum. We estimated the cutoff energies by calculating $E_{\text{cutoff}}=I_\text{p}+3.17U_\text{p}$, as given in the semiclassical rescattering model\cite{cork93prl,lewe+pra94}.

We observe that the grid and B-spline HHG spectra are indistinguishable for all the laser intensities. This fact is consistent with the analysis reported above on the spectrum of $\hat{H}_0$ (see Section \ref{field-freeH}). On the other hand, the agreement between the spectra obtained with the Gaussian basis and those obtained with the grid or B-splines deteriorates when the laser intensity increases. This is clearly observed for the plateau region for the intensity $I=5\times10^{14}$~W/cm$^2$, but also detected for the plateau and cutoff regions for the intensity $I=7\times10^{14}$~W/cm$^2$ (see Supplementary Information).  
Most of these observations are also valid when using the acceleration form of the HHG spectrum. The only exception we found was with the Gaussian basis set and laser intensities $I=5\times10^{14}$ W/cm$^2$, as shown in \reff{hhgdipacc5d14}, and $I=7\times10^{14}$ W/cm$^2$ (see Supplementary Information). For these largest intensities, the spectrum extracted from the acceleration seems to largely underestimate the position of the cutoff but to much better reproduce the harmonics of the plateau.

\begin{figure}[h!]
\centering
\includegraphics[scale=0.7]{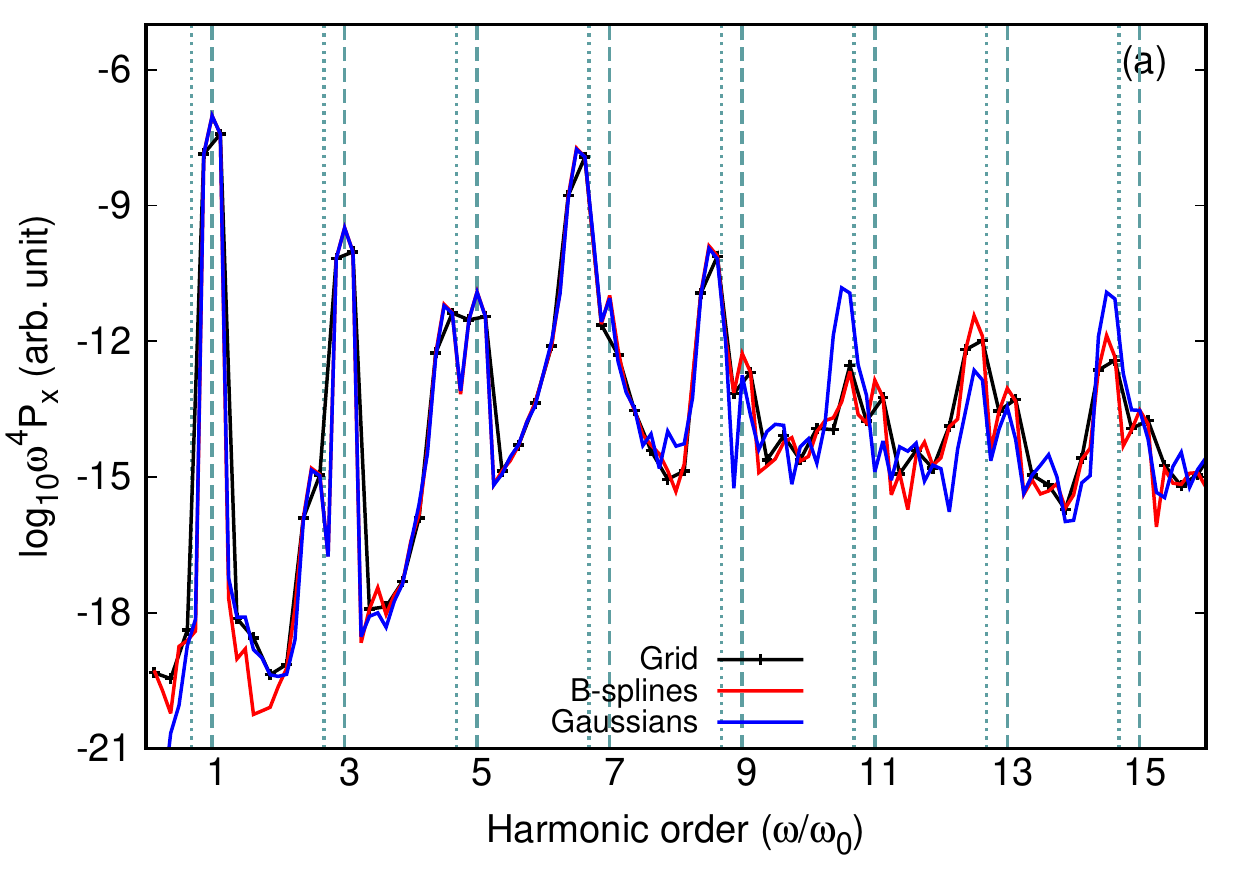} 
\includegraphics[scale=0.7]{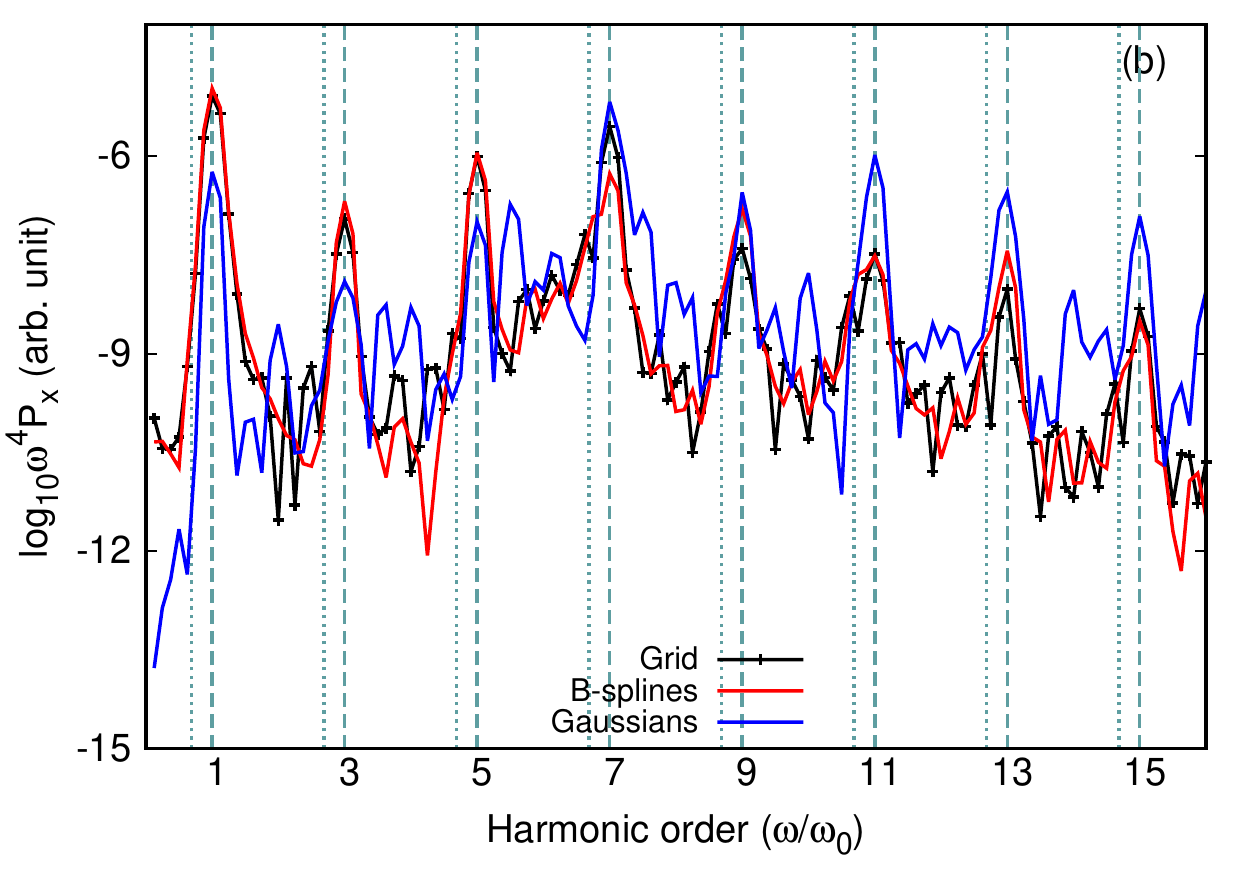} 
\caption{HHG spectra calculated from the electron dipole at the equilibrium internuclear distance $R$ = 2.0 au up to the 15th harmonic with laser intensities: (a) $I=10^{14}$~W/cm$^2$ and (b) $I=5 \times 10^{14}$~W/cm$^2$. The dashed lines indicate the position of the harmonics while the dotted lines indicate the hyper-Raman lines at position $\tilde{\omega} \pm 2 k \omega_0$ \cite{gau95} where $k$ is an integer and $\tilde{\omega}=6.69\omega_0$ is the resonance with the first excited state.}
\label{hhgdippics}
\end{figure}

To analyse in more details the fine structure of the HHG peaks, in \reff{hhgdippics} HHG spectra only up to the 15th harmonics. The B-spline and the grid spectra are almost identical except for some very small differences when the laser intensity is very high. Gaussian spectra reproduces the features of the B-spline and grid ones, but when the laser intensity increases the Gaussian spectrum become much more noisy.

From panel (a) of \reff{hhgdippics} it is also possible to identify another series of peaks besides those corresponding to the harmonics. These peaks corresponds to hyper-Raman lines with position given by $\tilde{\omega} \pm 2 k \omega_0$\cite{gau95}, where $k$ is an integer and $\tilde{\omega}=6.69\omega_0$ is the resonance with the first excited state. We observe that the three basis sets describe with the same accuracy the hyper-Raman lines. Moreover, at sufficiently large laser intensity, the HHG process dominates, and the hyper-Raman lines are not observed anymore (panel (b) of \reff{hhgdippics}).

The accuracy of the grid, B-spline, and Gaussian calculations was also investigated through their ability to reproduce the two-center interference in the HHG spectrum. This interference was predicted by Lein \emph{et al.}\cite{PhysRevA.66.023805} for diatomic molecules such as H$^{+}_{2}$. In this model, the electron that recombines with the ionic core can interact with either of the two nuclei. The two atomic centers can therefore be interpreted as coherent point sources and the whole system can be seen as a microscopic analog of Young's two-slit experiment. The light emitted by each nucleus will interfere either constructively or destructively depending on its frequency and the interference pattern will superimpose to the HHG spectrum. Since Lein's model has been proposed, a great number of numerical analyses came forth pointing out the role of the internuclear distance, molecular orientation, recombination to excited states, and laser intensity\cite{wor10,PhysRevA.76.043419,PhysRevA.87.043404,PhysRevA.95.033415,PhysRevA.73.023410,PhysRevA.77.013410,itat04nat,PhysRevLett.95.153902,PhysRevA.80.041403,Smi09}.

\begin{figure}[t]
\includegraphics[scale=0.7]{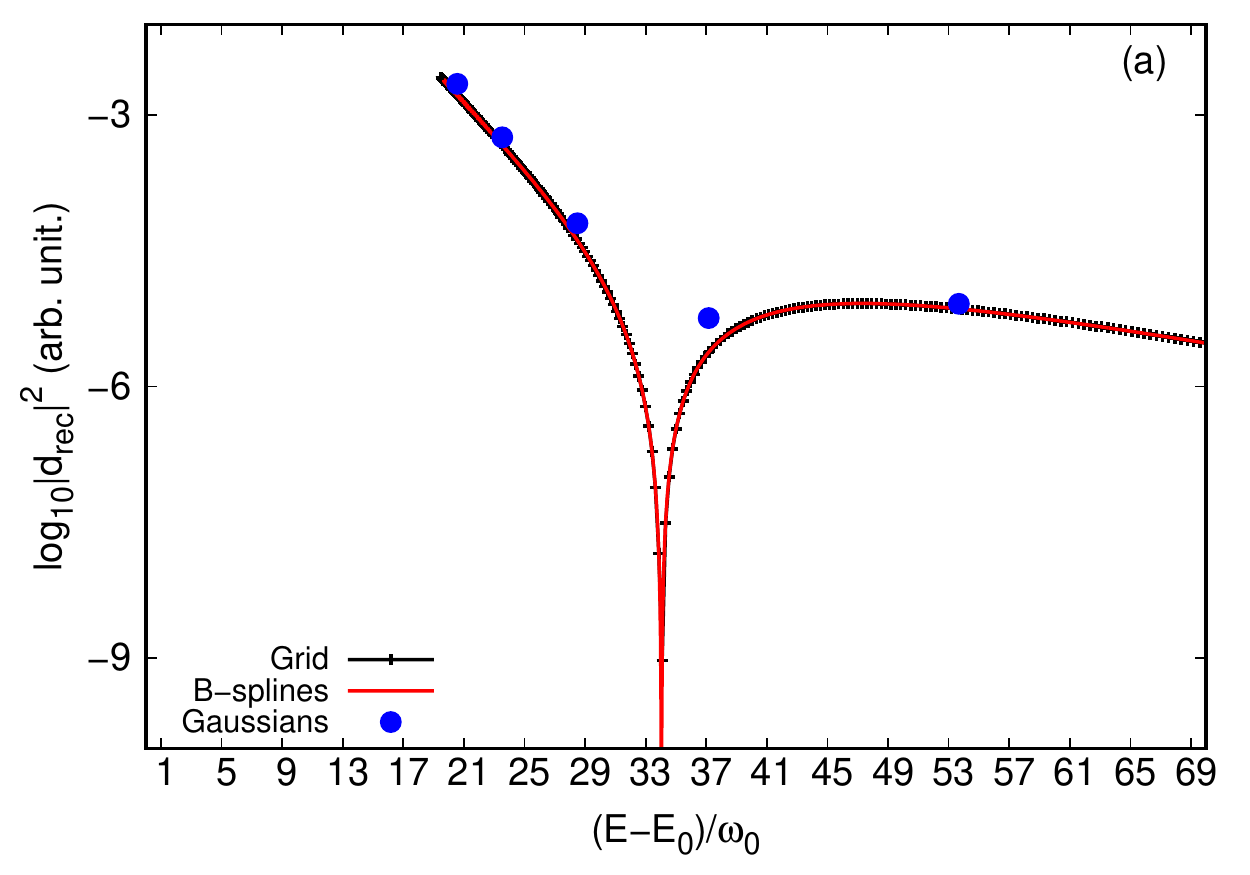} 
\includegraphics[scale=0.7]{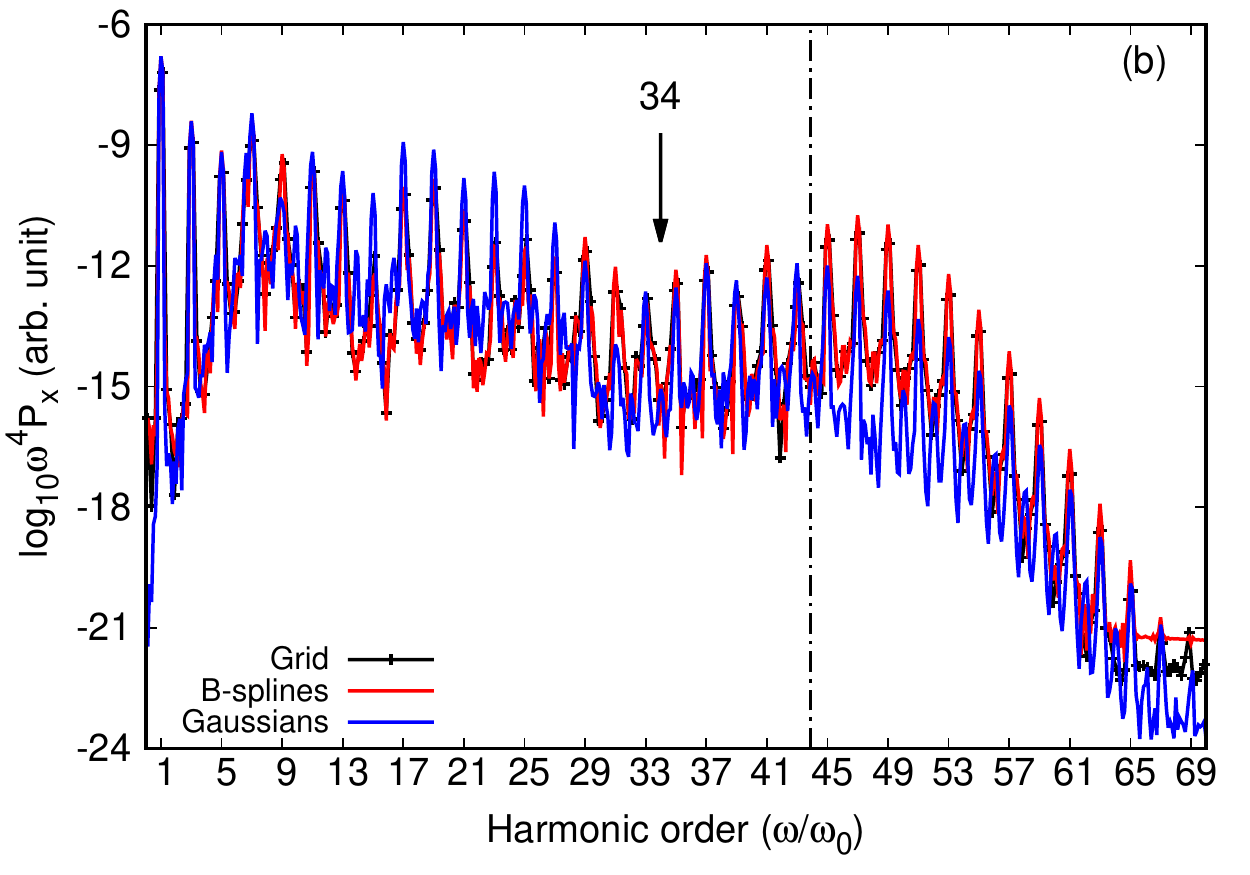}
\caption{Two-center interference at $R$ = 1.8 au: (a) recombination dipole and (b) HHG spectrum at $I=2 \times 10^{14}$~W/cm$^2$. The arrow points to the expected position of the two-center interference minimum extracted from the recombination dipole. The dot-dashed line is the cutoff energy $E_{\text{cutoff}} = 43.8\omega_0$. $E_{0}$ is the ground-state energy. }
\label{2centerR1.8}
\end{figure}

According to Lein's model, the position of the minimum in the spectrum is independent from the laser intensity and can be extracted from the analysis of the recombination dipole $d_{\text{rec}}(E)= \langle \varphi_0 | \hat{x} | \varphi_E \rangle$ where $\varphi_0$ is the ground state and $\varphi_E$ is a continuum state at energy $E$ of $\hat{H}_0$. This quantity is plotted in panel (a) of \reff{2centerR1.8} for $R$ = 1.8 au and in panel (a) of \reff{2centerR2.2} for $R$ = 2.2 au. For $R$ = 2.0 au, we report the recombination dipole in the Supplementary Information. The minimum described in the two-center interference corresponds to the energy which makes the recombination dipole vanishing. We found that the corresponding frequency is $\omega=34.0\omega_0$ for $R=1.8$ au, $\omega=26.4\omega_0$ for $R=2.0$ au, and $\omega=20.8\omega_0$ for $R=2.2$ au. We note that the extraction of the minimum from the recombination dipole is straightforward for the grid and B-spline basis sets, while in the case of the Gaussian basis only a rough estimate can be given. Lein's model predicts the position of the minimum at $\omega = {\pi^2}/{(2R^2\omega_0)}$ which gives $\omega=26.7\omega_0$ for $R=1.8$ au, $\omega=21.6\omega_0$ for $R=2.0$ au, and $\omega=17.9\omega_0$ for $R=2.2$ au. The underestimation of the minimum position by Lein's model has already been pointed out\cite{PhysRevA.73.023410}. The main reasons must be searched in the different description of the ground state and the continuum between our 1D theoretical model and Lein's model.

We report in panel (b) of \reff{2centerR1.8} and in panel (b) of \reff{2centerR2.2} the HHG spectra for $R=1.8$~au and for $R=2.2$~au with $I=2\times10^{14}$~W/cm$^2$ and we observe that all the basis sets reproduce the position of the minimum of the two-center interference. Also the minimum for $R=2.0$~au is very well reproduced as can be seen in \reff{hhgdip}. Another observation is that the sharpness of the minimum depends on the laser intensity and on the internuclear distance. We confirm the fact that the minimum is more visible for smaller internuclear distances\cite{RisoudSciRep}. We did the same investigation considering the recombination acceleration $a_{\text{rec}}(E)= \langle \varphi_0 | -\nabla \hat{V} | \varphi_E \rangle$ and the HHG spectrum from the acceleration. We obtained the same results (see Supplementary Information) explained before. From these studies we deduce that all the basis sets are capable to accurately reproduce the two-center interference\cite{PhysRevA.66.023805}. However, in the case of the Gaussian basis, the acceleration seems to better reproduce the minimum for $I=5 \times 10^{14}$~W/cm$^2$ (panel (c) of \reff{hhgdipacc5d14}) and $I=7 \times 10^{14}$~W/cm$^2$ (see Supplementary Information).

\begin{figure}
\centering
\includegraphics[width=0.5\textwidth]{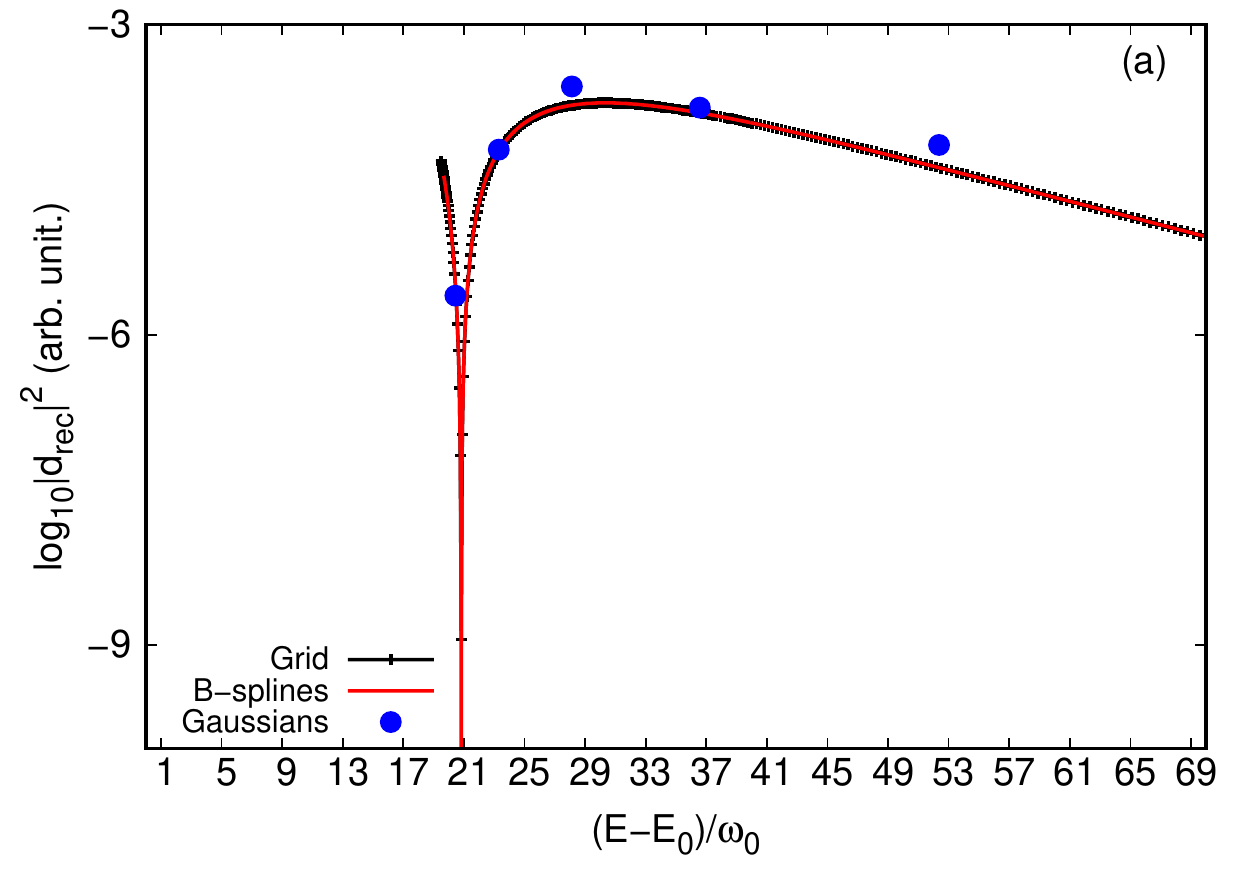}
\includegraphics[width=0.5\textwidth]{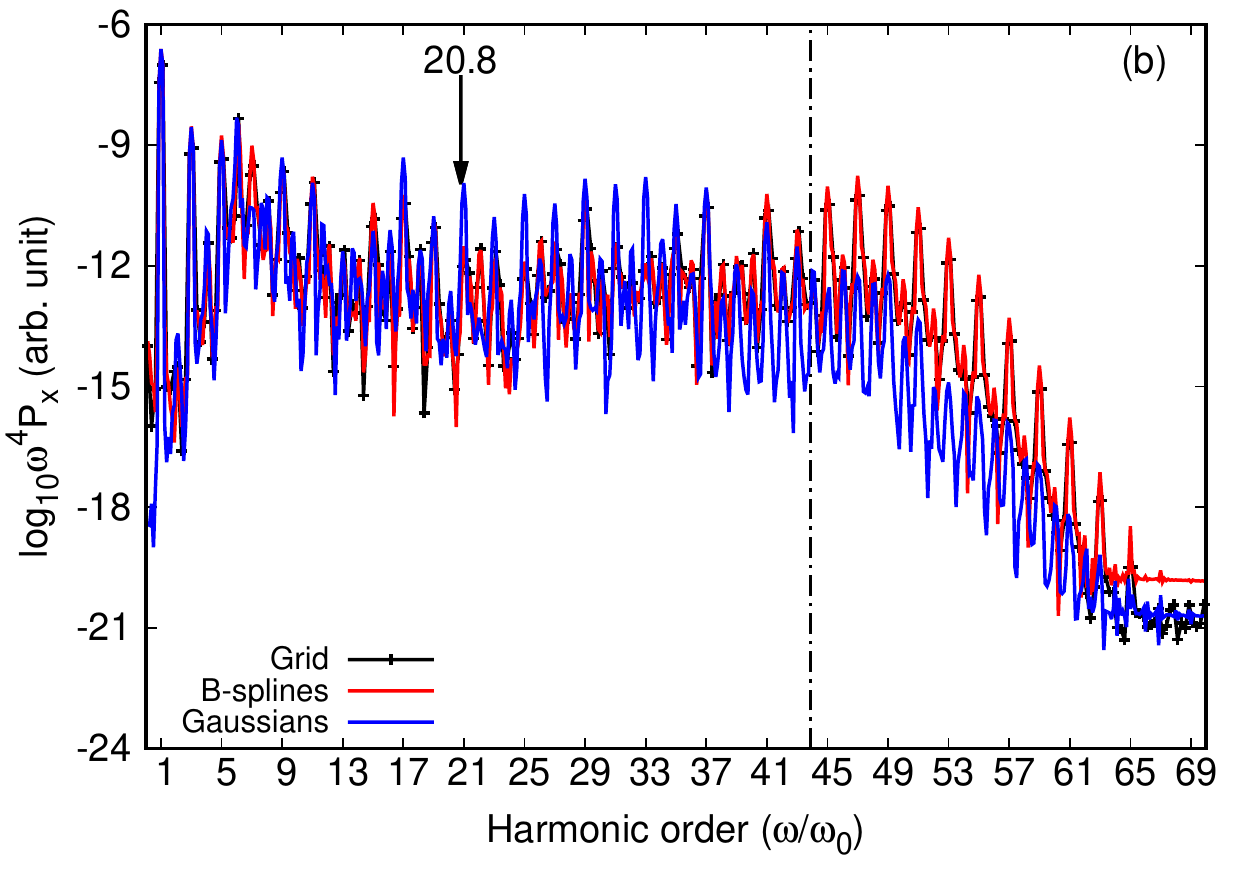}
\caption{Two-center interference at $R$ = 2.2 au: (a) recombination dipole and (b) HHG spectrum at $I=2 \times 10^{14}$~W/cm$^2$. The arrow points to the expected position of the two-center interference minimum extracted from the recombination dipole. The dot-dashed line is the cutoff energy $E_{\text{cutoff}} = 43.8\omega_0$.  $E_{0}$ is the ground-state energy.}
\label{2centerR2.2}
\end{figure}

\begin{figure}
\centering
\includegraphics[scale=0.7]{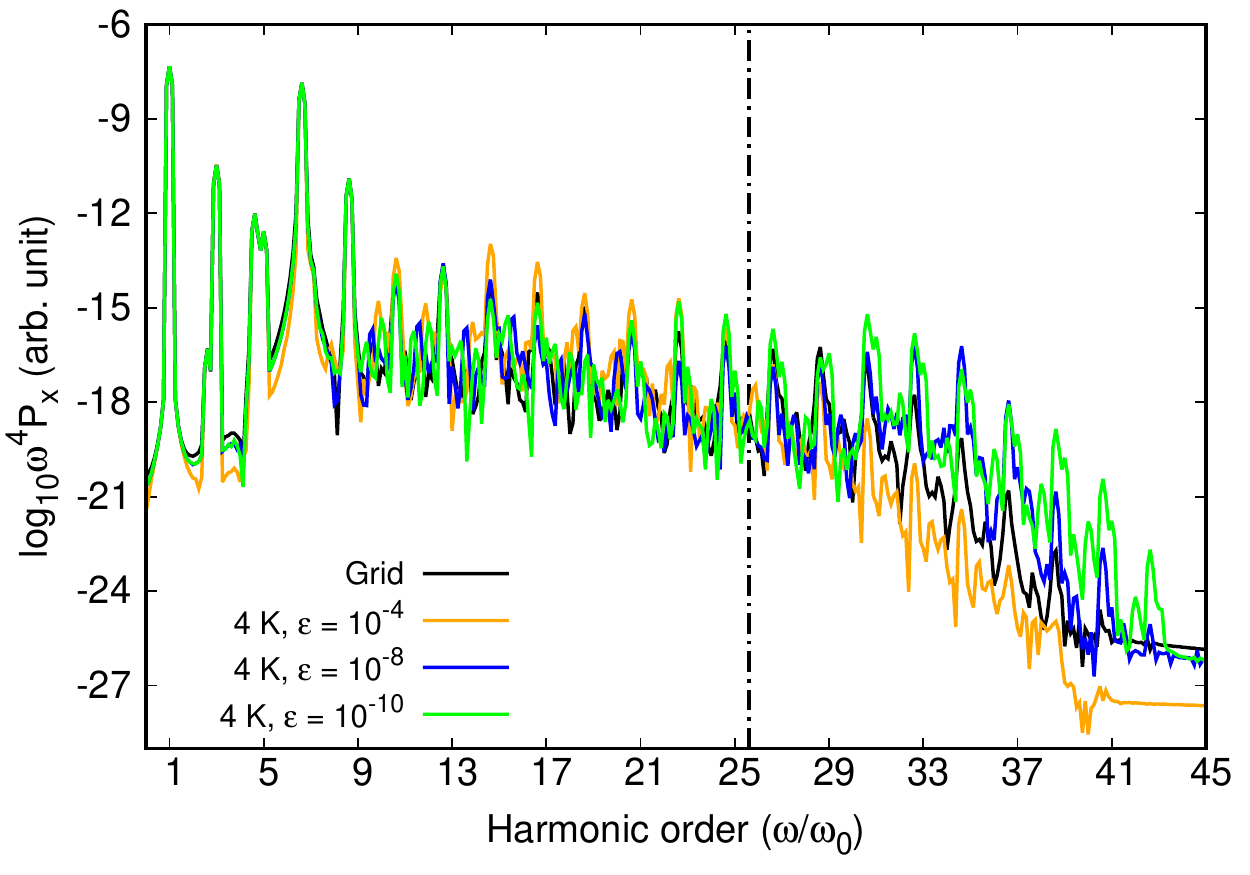}
\caption{HHG spectra from the dipole at the equilibrium internuclear distance $R$ = 2.0 au with $I=5\times10^{13}$ W/cm$^2$ obtained with the grid and with the Gaussian basis sets with linear-dependency thresholds $\epsilon=10^{-4}$, $\epsilon=10^{-8}$, and $\epsilon=10^{-10}$.}
\label{convergence}
\end{figure}

From the detailed analysis of HHG spectra presented in this section, we conclude that for a good performance of the Gaussian basis the laser intensity can not be ``very large''. For example, for intensity lower than $I=5\times10^{14}$~W/cm$^2$ we obtain correct HHG spectra while for higher intensities only the harmonic peaks in the low-energy part of the plateau are correct. A strategy to improve the Gaussian basis set could be to modify the cutoff $\epsilon$ below which the eigenvalues of the overlap matrix are set to zero. This will change the number of kept eigenvectors. In \reff{convergence} we compare an HHG spectrum for $I=5\times10^{13}$ W/cm$^2$ calculated with the grid and with the Gaussian basis while changing the linear-dependency threshold $\epsilon$: $\epsilon=10^{-4}$ (17 basis functions), $\epsilon=10^{-8}$ (24 basis functions, which is the standard choice throughout the article), and $\epsilon=10^{-10}$ (26 basis functions).
This analysis shows that for  a ``low'' intensity ($I=5\times10^{13}$ W/cm$^2$) the quality of the HHG spectrum in the plateau and cutoff regions is not affected by the specific choice of the threshold of eigenvalues.

\subsection{ATI}

\begin{figure}[h!]
\centering
\includegraphics[scale=0.7]{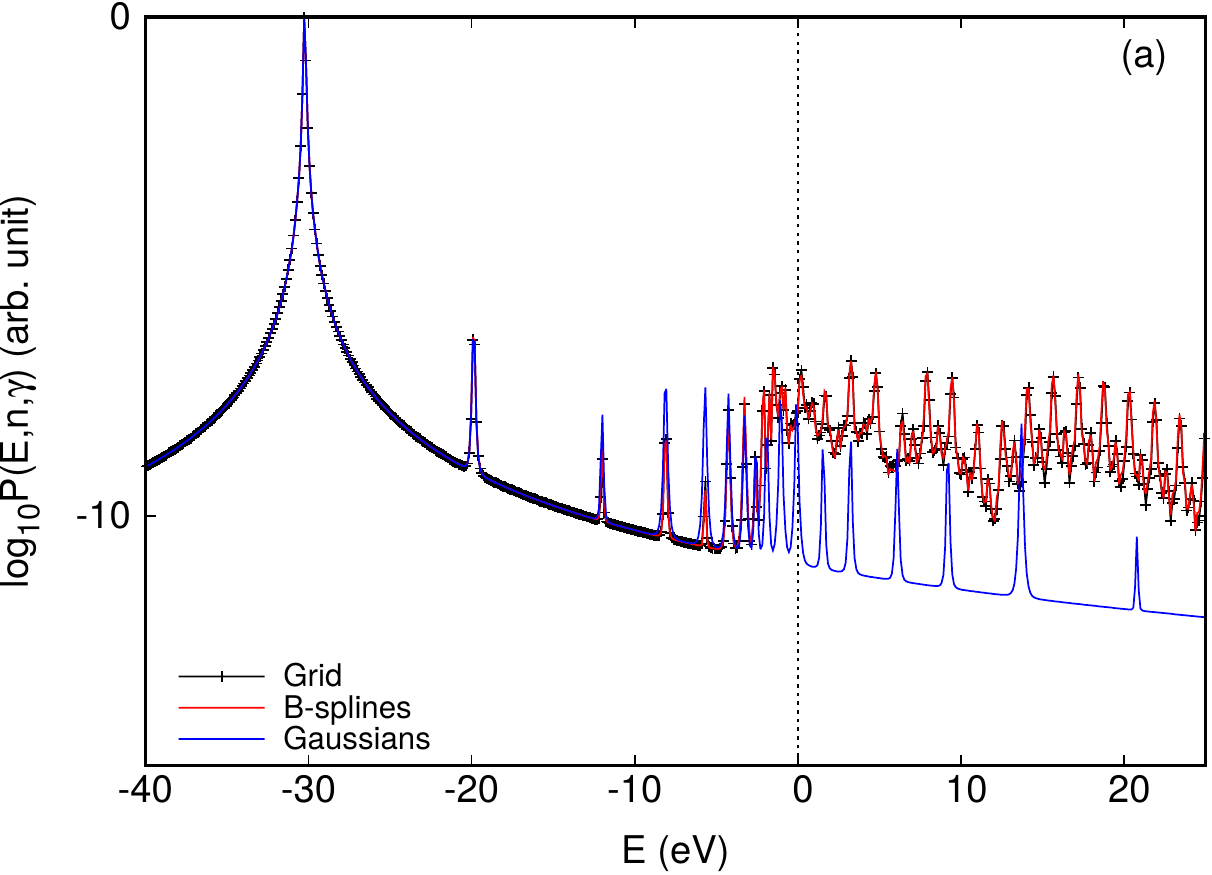}\\
\includegraphics[scale=0.7]{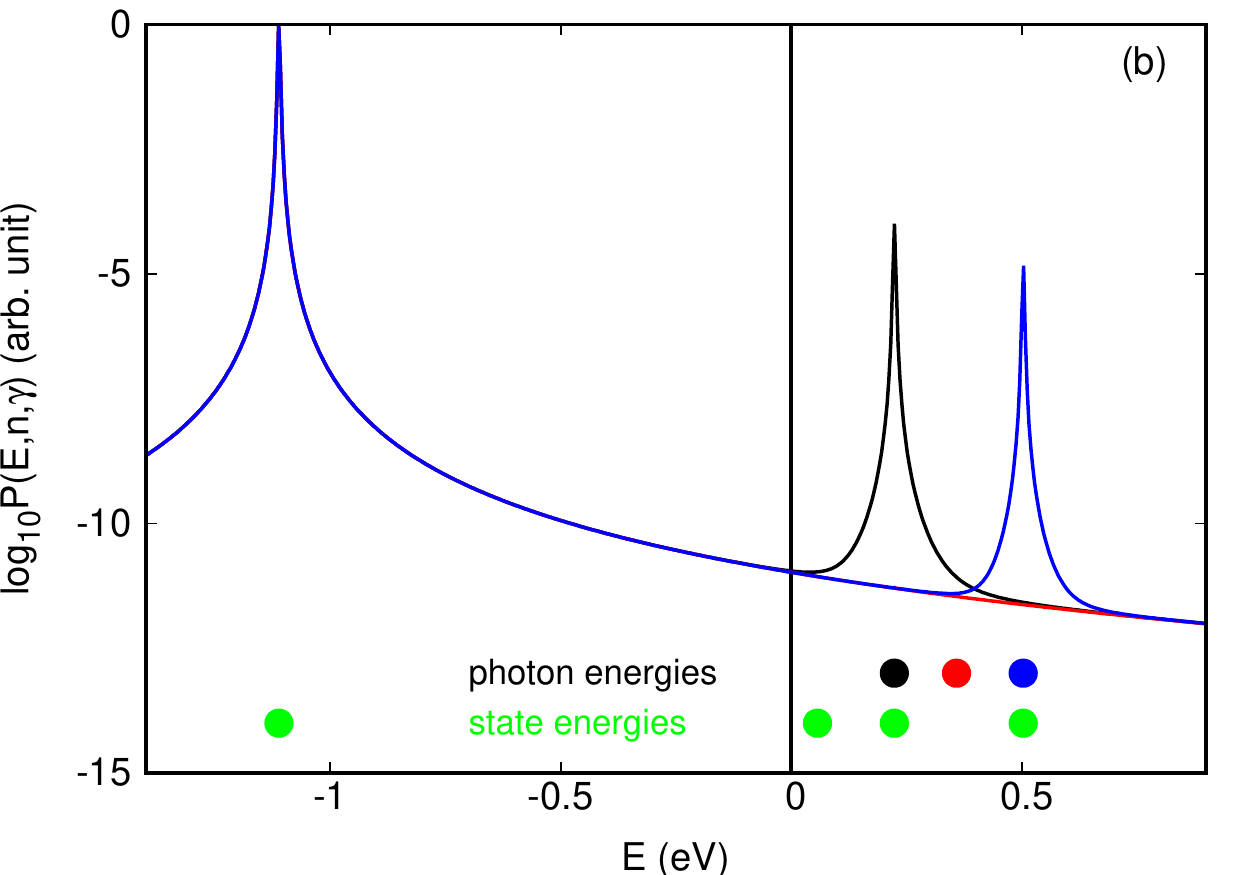}
\caption{{(a)} ATI spectrum calculated at the equilibrium interatomic distance $R$ = 2.0 au with intensity $I=1\times 10^{14}$~W/cm$^2$. {(b)} Photoelectron spectrum calculated with the Gaussian basis at the equilibrium distance $R$ = 2.0 au with intensity $I=1 \times 10^{14}$~W/cm$^2$ and three photon energies $\omega_0$ = 1.34 Ha (black), $\omega_0$ = 1.47 Ha (red), and $\omega_0$ = 1.61 Ha (blue). The ground-state energy (-1.11 Ha) and the continuum-state energies (0.06 Ha, 0.22 Ha, and 0.50 Ha) which correspond to transitions allowed by symmetry are displayed (magenta dots).}
\label{ati}
\end{figure}

We calculated ATI spectra with intensities $I=5\times10^{13}$, $1\times10^{14}$, and $5\times10^{14}$~W/cm$^2$. In panel (a) of \reff{ati} we show the ATI spectrum with laser intensity $I=10^{14}$~W/cm$^2$, while the spectra for intensities $I=5\times10^{13}$ and $5\times10^{14}$~W/cm$^2$ are reported in the Supplementary Information. 

The ATI spectrum of \reff{ati} has positive energy peaks (bound-continuum transitions) corresponding to the electron density ionized during the propagation, i.e. the photoelectron spectrum, while the peaks in the negative region  (bound-bound transitions) represent the electron density remaining in the ground state and that has been transferred to excited states. We remind that only the positive energy region of an ATI spectrum is experimentally measurable.

As already seen for the HHG spectra, the grid and B-spline basis sets describe with the same accuracy both bound-bound and bound-continuum transitions. Their ATI spectra coincide and correctly reproduce the expected features of an ATI spectrum: the distance between two consecutive ATI peaks (in the positive energy region) is constant and equal to the energy of a photon, \text{i.e.} $0.057$ Ha.

The Gaussian basis is only able to reproduce bound-bound transitions. The negative energy part of the spectrum is quite close to the one obtained with the grid and B-splines, while bound-continuum transitions are out of reach for the Gaussian basis set. This limitation is due to the low density of states in the continuum. Indeed, with the basis-set parameters used here, only six continuum states are reproduced in the energy region between 0 and 1 Ha, as we can see in the bottom panel of \reff{DOS}.  This low density of states is far from reproducing the correct ATI energy distribution and explains why no more than six peaks are observed in the positive energy region of the spectrum. The energies of the six ATI peaks correspond to the energies of the six continuum states reported in \reff{DOS}. 
To detail more on this feature, we plot in panel (b) of \reff{ati} the photoelectron spectrum, computed with the Gaussian basis, after absorption of one photon and for three different photon energies $\omega_0$ = 1.34 Ha, $\omega_0$ = 1.47 Ha, and $\omega_0$ = 1.61 Ha. Together, we also plot the energy position of the ground state and of the first continuum energies corresponding to symmetry-allowed transitions. One clearly sees that if the photon energy matches the energy of a transition from the ground state to one of the continuum states then we get a photoelectron peak.  However, if the photon energy does not match any transition then no ionization is observed. This crucial feature forbids the computation of a correct photoelectron or ATI spectrum with the Gaussians basis set used here. We believe that larger Gaussian basis sets can in principle describe ATI. Indeed, in 3D calculations\cite{coccia16b}, one can easily produce tens of low-energy ($<$1 Ha) continuum states, leading to a possible improvement of the ATI spectrum.

\section{3D theoretical model of H$^{+}_{2}$}
\label{theory3D}
\noindent

The electronic TDSE for a 3D model of H$^{+}_{2}$ is given by, in atomic units (au),
\begin{equation}
\label{TDSE3D}
i\frac{\partial}{\partial t}\psi({\bf r},t) = \left[\hat{H}_{0}({\bf r}) + \hat{H}_\mathrm{int}({\bf r},t) \right] \psi({\bf r},t),
\end{equation}
where $\psi({\bf r},t)$ is the time-dependent electron wave function. Here, $\hat{H}_0({\bf r})$ is the field-free Hamiltonian,
\begin{equation}
\label{free}
\hat{H}_0({\bf r})= -\frac{1}{2} \nabla^2 + \hat{V}({\bf r}),
\end{equation}
with $\hat{V}({\bf r})$ the Coulomb electron-nuclei interaction. 

The interaction between the electron and the laser electric field $E(t)$ is taken into account by the time-dependent interaction potential, which is given in the length gauge by
\begin{equation}
\hat{H}_\mathrm{int}({\bf r},t)=\hat{z}E(t),
\end{equation}
where $E(t)$ is the laser electric field polarized along the $z$ axis, corresponding to the H$_{2}^{+}$ internuclear axis, and $\hat{z}$ is the electron position operator along this axis. We have chosen the same type of laser as in the 1D model (see Section \ref{theory1D}) except that the duration of the pulse is $\tau = 6 T_0$ (i.e., 6 optical cycles). We calculated HHG spectra from the dipole as in \refe{HHGacc}.

\subsection{Representation of the time-dependent wave function and propagation}

\begin{figure}[ht!]
\includegraphics[scale=0.67]{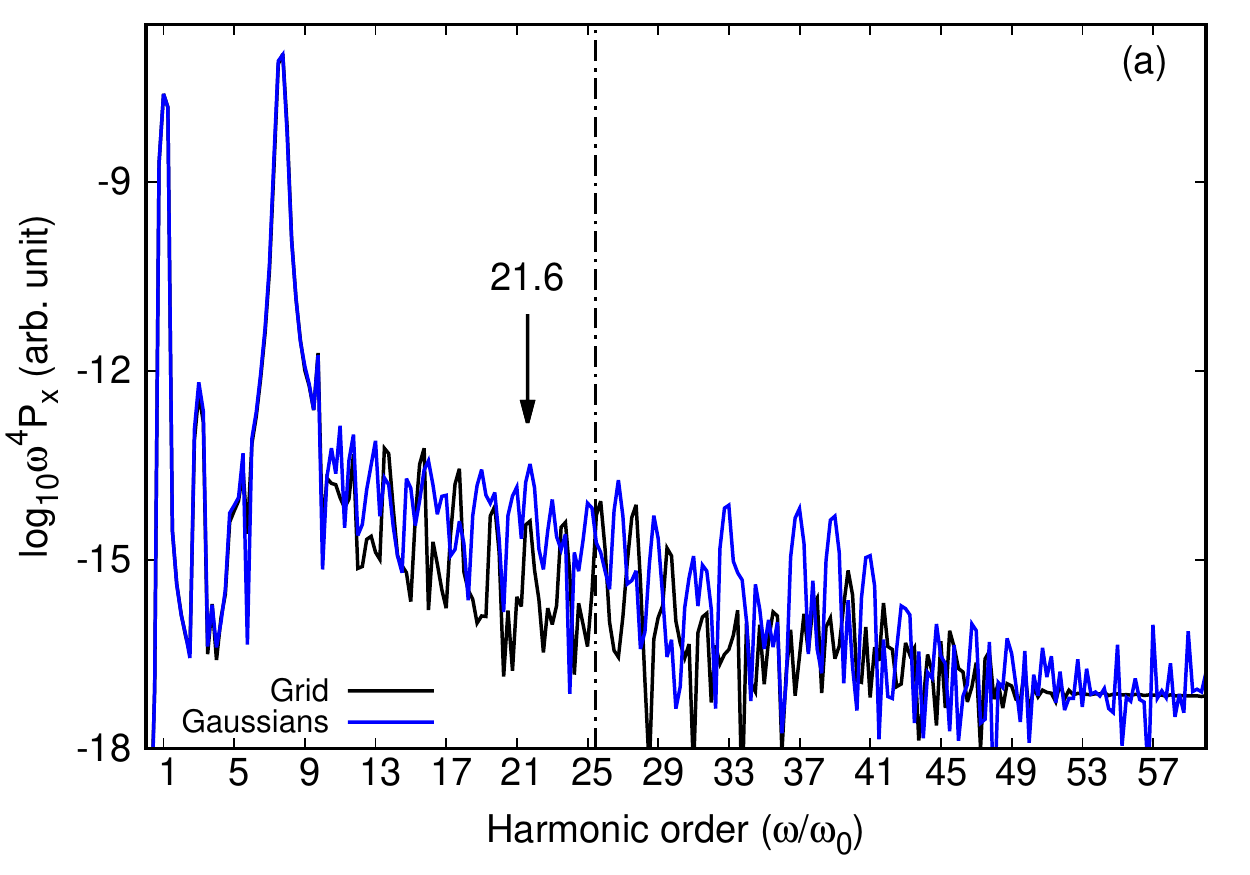}
\includegraphics[scale=0.67]{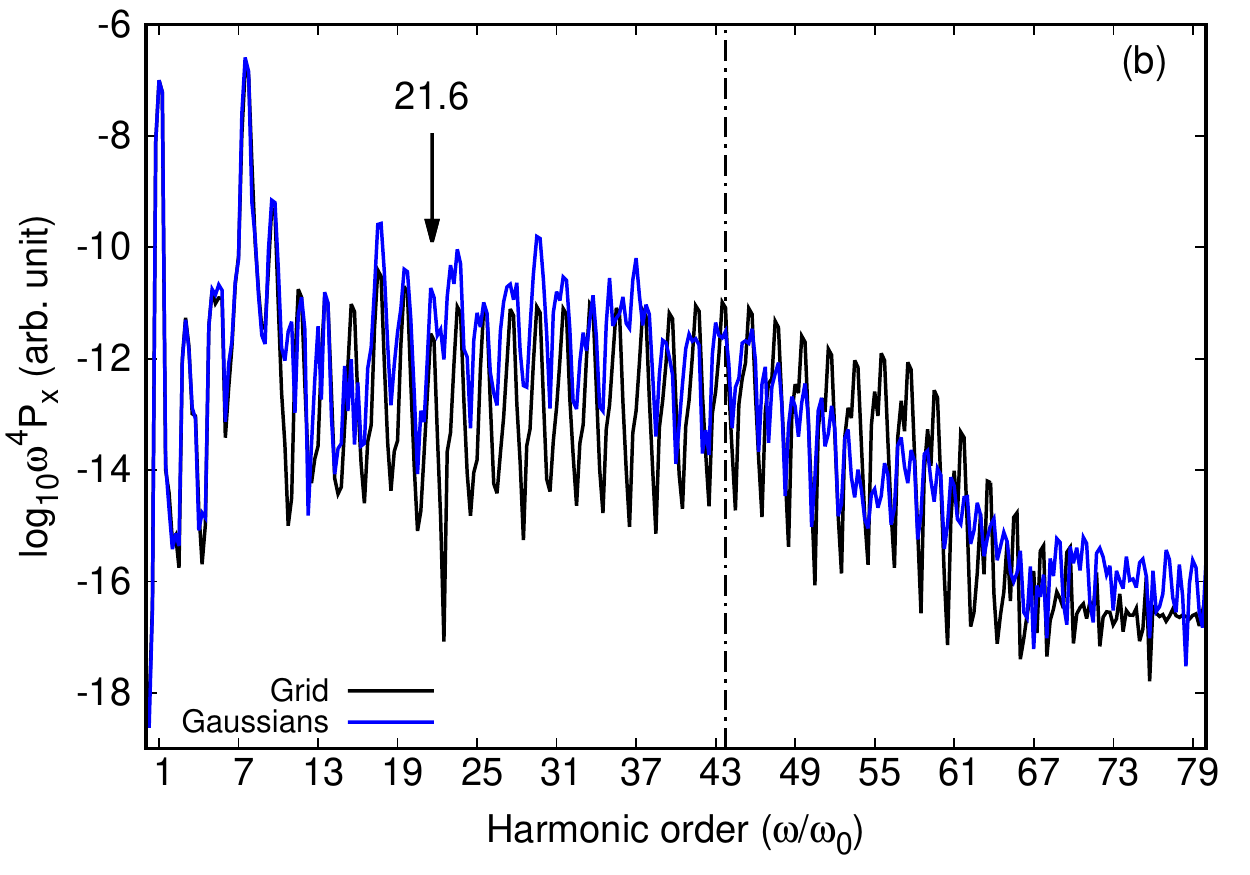}
\includegraphics[scale=0.67]{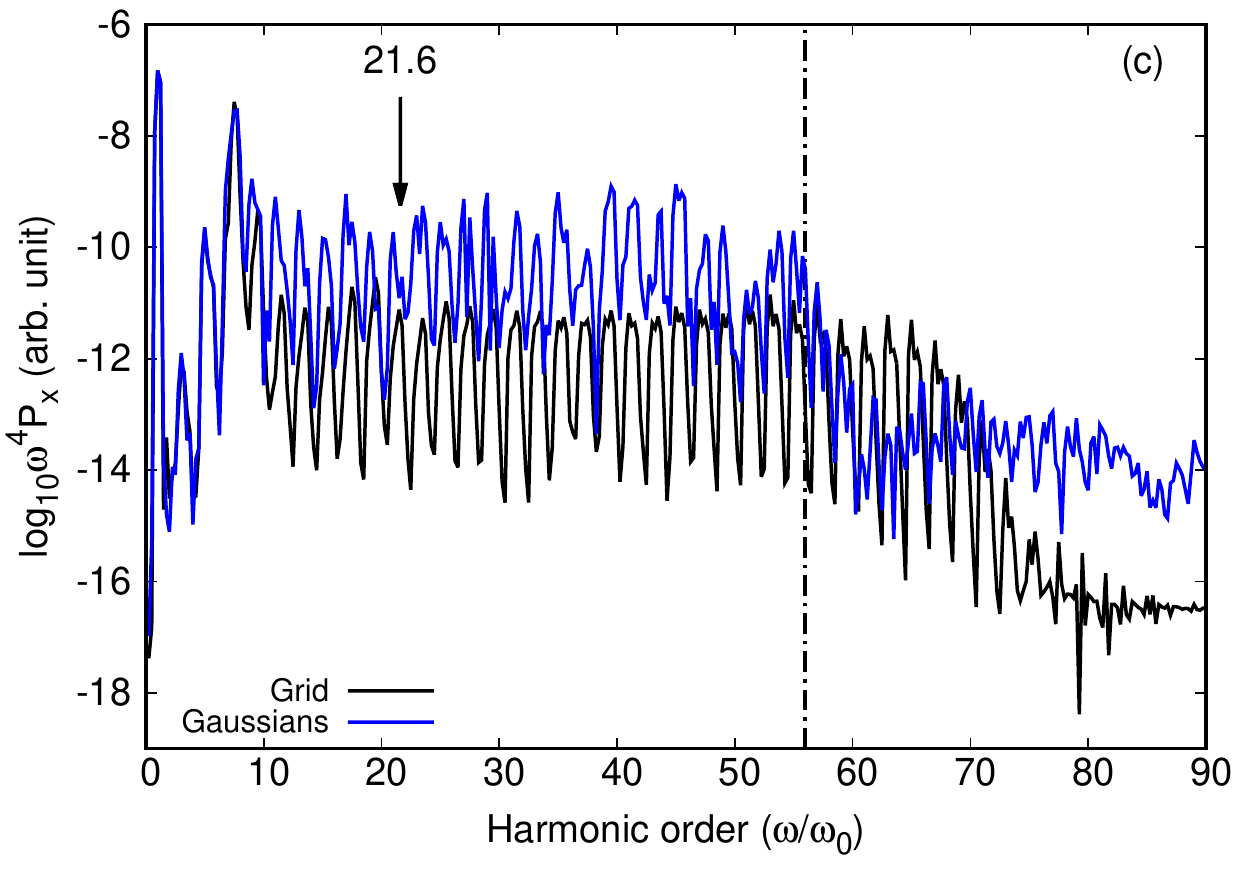}
\caption{HHG spectra in the dipole form at the equilibrium internuclear distance $R$ = 2.0 au with laser intensities: (a) $I=5 \times 10^{13}$ W/cm$^2$, (b) $I=2 \times 10^{14}$ W/cm$^2$, and (c) $I=3\times 10^{14}$ W/cm$^2$. For each HHG spectrum, the dot-dashed line gives the cutoff energy $E_{\text{cutoff}}=I_\text{p}+3.17U_\text{p}$ given by the rescattering model\cite{cork93prl,lewe+pra94} which is (a) $E_{\text{cutoff}} = 25.4\omega_0$, (b) $E_{\text{cutoff}} = 43.7\omega_0$, and (c) $E_{\text{cutoff}} = 55.9\omega_0$. The arrow points to the expected position of the two-center interference minimum extracted from the recombination dipole.}
\label{3Dhhgdip}
\end{figure}

\begin{figure}[ht!]
\includegraphics[scale=0.7]{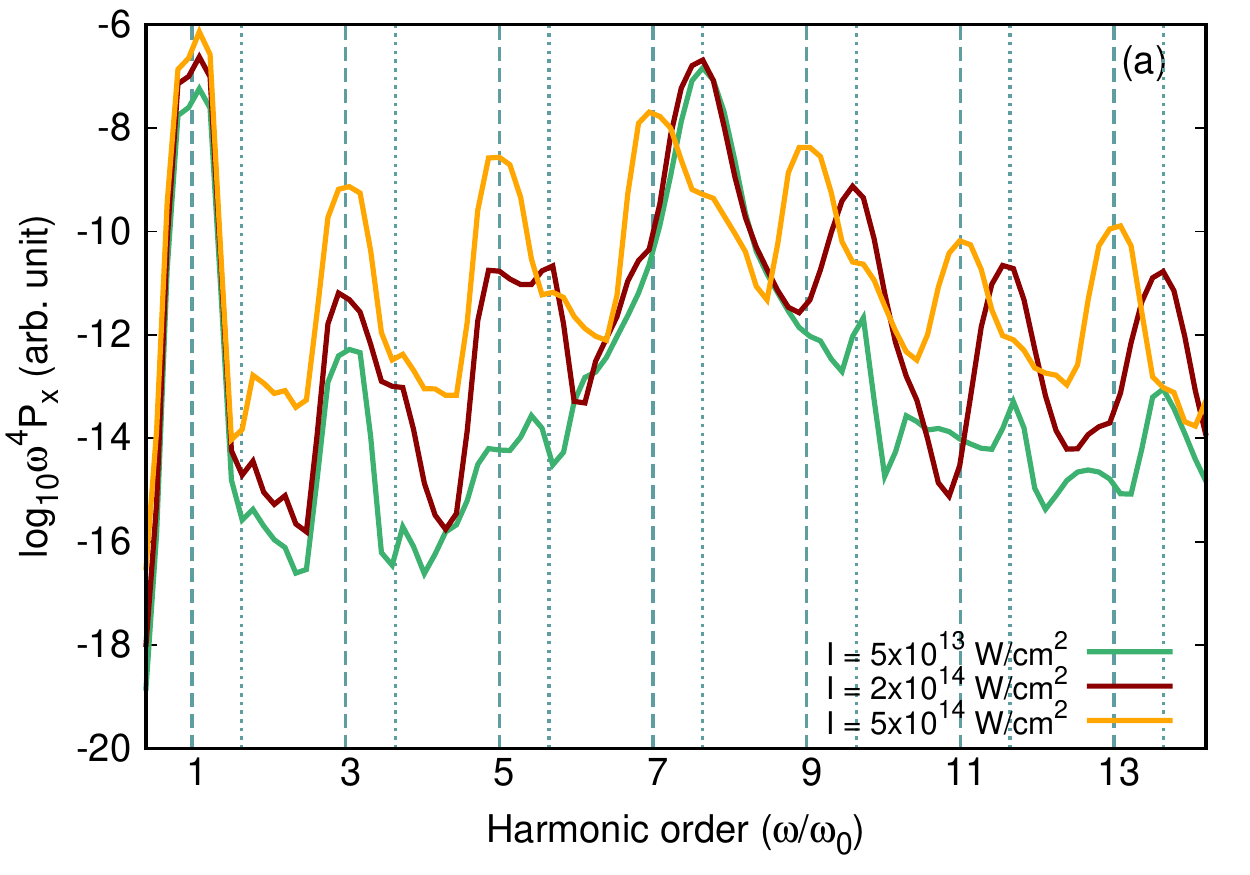}
\includegraphics[scale=0.7]{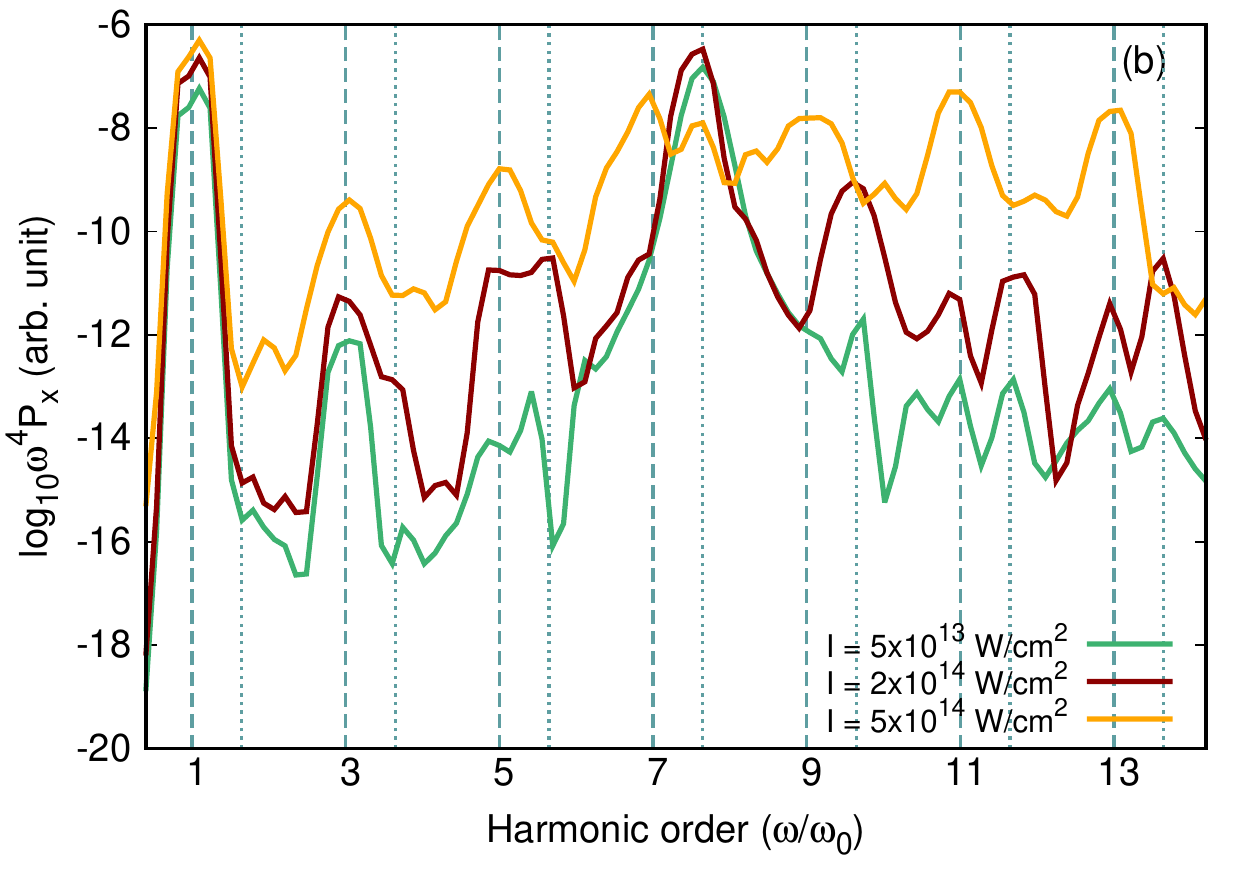}
\caption{HHG spectra in the dipole form at the equilibrium internuclear distance of $R=2.0$ au up to the 13th harmonic with laser intensities :  $I=5\times10^{13}$ W/cm$^2$, $I=2\times10^{14}$ W/cm$^2$, and $I=5\times10^{14}$ W/cm$^2$ for (a) grid and (b) Gaussian basis sets. For each HHG spectrum, the dashed line indicates the position of the harmonics and the dotted line indicates the hyper-Raman lines at position $\tilde{\omega} \pm 2 k \omega_0$ \cite{gau95} where $k$ is an integer and $\tilde{\omega}=7.65\omega_0$ is the resonance of the first excited state.}
\label{3Dhhgdippics}
\end{figure}

\subsubsection{Real-space grid} 

Concerning the 3D calculations on a grid, we used the Octopus code which is a software package for TDDFT calculations.\cite{Octopus2015} For our calculations we have chosen the ``independent particle'' option which permits to get the numerically exact solution for the TDSE in the case of one electron. We have chosen as simulation box a cylinder with radius 50 au and height 100 au with a grid space $\Delta r$ = 0.435 au. The TDSE of \refe{TDSE3D} is solved by means of the Crank-Nicholson propagation algorithm\cite{crank_nicolson,numerical_recipies} with a time step $\Delta t = 5\times10^{-2}$ au. 
Also in this case to avoid unphysical reflections at the boundaries of the simulation box, a mask-type absorber function was used with a spatial extension of 22 au. 

\subsubsection{Gaussian basis set}

In this case, we used the approach we developed and detailed in Ref.\cite{lupp+13jcp,coccia16b} which consists in solving the TDSE using the TDCI approach. For the Gaussian calculations, we used a development version of the MOLPRO software package\cite{Molpro-PROG-15} and the external code LIGHT\cite{lupp+13jcp} to perform the time propagation using also in this case a time step $\Delta t=5\times 10^{-2}$ au. As Gaussian basis set we used a 6-aug-cc-pVTZ with 5 K functions, which we denote as 6-aug-cc-pVTZ+5K, which is the largest basis without linear dependencies.  The basis-set exponents and contraction coefficients are collected in Table S2 of Supporting Information. To treat ionization we used a double-$d$ heuristic model where the parameters $d_1$ and $d_0$ have been chosen as in the 1D model. The value of $I_\mathrm{p}$ is in this case -1.10 Ha.

\section{3D RESULTS AND DISCUSSION \label{results3D}}

\subsection{HHG}

We calculated HHG spectra in the dipole form for H$_{2}^{+}$ at internuclear distance $R$ = 2.0 au (equilibrium) for a Ti:Sapphire laser with a carrier frequency $\omega_0=0.057$ Ha and intensities $I=5\times10^{13}$, $1\times10^{14}$, $2\times10^{14}$, $3\times10^{14}$, $4\times 10^{14}$, and $5\times10^{14}$ W/cm$^2$.

In \reff{3Dhhgdip} we show the HHG spectra for three laser intensities (the spectra for the other intensities are reported in the Supplementary Information). Both the Gaussian and grid basis sets reproduce well the expected features of an HHG spectrum, regardless of the applied field intensity, as already pointed out for the 1D case. However, starting from intensity $I=3\times10^{14}$ W/cm$^2$, the quality of the spectrum obtained with the Gaussian basis set tends to diminish, especially in the cutoff region. For 3D calculations, obtaining a good HHG spectrum with optimized Gaussians seems to be more difficult than for 1D calculations, due to the computational complexity.

However, it is interesting to note that the low-energy harmonics are still well described when compared to the grid calculations. We show this behavior by analysing the fine structures of the peaks as shown in \reff{3Dhhgdippics}. Here, we plot the HHG spectra up to the 13th harmonic for different intensities. For the grid calculations (panel (a)) with $I=5\times10^{13}$ W/cm$^2$ only the first and the third harmonic peaks are clearly visible together with a strong and large peak at around 7.65$\omega_0$, due to the emission from the first excited state. Also in this case we observe hyper-Raman lines at position $\tilde{\omega} \pm 2 k \omega_0$ \cite{gau95} where $k$ is an integer and $\tilde{\omega}=7.65\omega_0$ is the resonance with the first excited state. Observing the evolution of the harmonics and the resonant peaks as a function of the laser intensity (from $I=5\times10^{13}$ W/cm$^2$ to $I=5\times10^{14}$ W/cm$^2$), the harmonics become more and more intense while the hyper-Raman lines almost disappear. The same behaviour was already observed in the 1D model. The spectra obtained with the Gaussian basis set show exactly the same trend as shown in panel (b) of \reff{3Dhhgdippics}.

\section{CONCLUSIONS}
\label{conclusions}

We explicitly solved the 1D and 3D TDSE for H$^{+}_{2}$ in the presence of an intense electric field and we explored the numerical performance of using a real-space grid, a B-spline basis, or a Gaussian basis optimized for the continuum. We analyzed the performance of the three basis sets for calculating HHG and ATI spectra. In particular, for HHG, the capability of the basis set to reproduce the two-center interference and the hyper-Raman lines was investigated. We showed that the grid and B-spline representations of the time-dependent wave function give the same results for both HHG and ATI. On the contrary, the performance of the Gaussian basis is more mixed and depends on the intensity of the laser. It is possible to optimize Gaussian functions to describe the low-energy part of the continuum. However, this optimization is limited by the issue of linear dependencies among Gaussian functions. This implies that for HHG the Gaussian basis can perform well up to the laser intensity $I=5\times10^{14}$ W/cm$^2$ for 1D and up to $I=2\times10^{14}$ W/cm$^2$ for 3D. For higher intensities we have found that only low-energy harmonics are still correct. Moreover, for 3D calculations, obtaining a good HHG spectrum with optimized Gaussian functions seems to be more difficult than in 1D calculations. Despite their limitations, Gaussian basis sets can reproduce intricate features of the HHG spectrum at low energy. Instead, in the case of ATI, Gaussian basis sets make impossible the description of a correct spectrum.  \\
In conclusion, from our investigation we noticed that the grid and B-spline basis sets have very similar behavior and computational cost. These basis sets are very accurate to describe the continuum and phenomena such as HHG and ATI. Gaussian basis sets are less efficient to describe the continuum.  The effect on ATI and HHG spectra is however different: on one hand, ATI spectrum is not  reproduced by Gaussian basis functions, on the other hand  the most important features and fine structures (minimum/resonances) at low energy of the HHG spectrum are correctly described. A clear advantage of Gaussian functions with respect the other basis sets is their computational cost which continues to make them interesting for many-electron systems.

\section*{Acknowledgement}

We acknowledge financial support from the LABEXs MiChem and PlasPar (ANR-11-IDEX-0004-02)  and the program ANR-15-CE30-0001-01-CIMBAAD. EC acknowledges funding from the ERC under the grant ERC-CoG-2015 No. 681285 "TAME-Plasmons".

\section*{Supporting Information}
Exponents and contraction coefficients of the Gaussian basis sets for the 1D (Table S1) and 3D (Table S2) cases;
dipole forms of HHG spectra from 1D calculations for $I=5 \times 10^{13}$ W/cm$^2$ (Figures S1 and S2) and  $I=7 \times 10^{14}$ W/cm$^2$ (Figures S3 and S4);  acceleration forms of HHG spectra  from 1D calculations with different intensities (Figures S5-S18); ATI spectra (Figures S19 and S20);  comparison between dipole and acceleration forms of HHG spectra at different intensities (Figures S21-S25); two-center interference analysis (Figures S26-S29); dipole forms of HHG spectra from 3D calculations at different intensities (Figures S30-S32); analysis of dipole forms of HHG and resonant peaks from 3D calculations at different intensities (Figure S33).

\bibliographystyle{acs}
\bibliography{bib}

\begin{mcitethebibliography}{78}
\bibitem[ati()]{ati}
 For the grid and B-spline basis sets, the ATI signal was evaluated following
  the approach explicitly given in Ref \cite{SCHAFER91}. We compute $
  P(E,n=2,\gamma) = \langle \psi(\tau) | \gamma^4/[(\hat{H}_{0}-E)^4 +
  \gamma^4] | \psi(\tau) \rangle = \gamma^4 \langle \chi | \chi \rangle$, where
  $|\chi \rangle$ is defined in Eq. (2) of Ref \cite{SCHAFER91}: $
  (\hat{H}_{0}-E+ \sqrt{i}\gamma)(\hat{H}_{0}-E-\sqrt{i}\gamma) \vert \chi
  \rangle = | \psi(\tau) \rangle$. Then $P(E,n=2,\gamma) = \gamma^4 \langle
  \chi | \chi \rangle$ is directly obtained from the norm of $ \vert \chi
  \rangle$. For the Gaussian basis sets, the wave function is expressed in a
  orbital basis for which the window operator of Eq. (\ref{PHOTO}) is diagonal.
  We thus simply evaluate $P(E,n=2,\gamma) = \sum_j |c_j|^2 \gamma^4/[
  (E_j-E)^4 + \gamma^4]$, where $j$ runs over the (discrete) states from the
  quantum chemistry calculation.\relax
\mciteBstWouldAddEndPunctfalse
\mciteSetBstMidEndSepPunct{\mcitedefaultmidpunct}
{}{\mcitedefaultseppunct}\relax
\EndOfBibitem
\bibitem[str()]{str}
 The Str\"omgren procedure consists in fitting the asymptotic form of the
  numerical solution of the Schr\"odinger equation [Eq. (1)] with the exact
  solution calculated when $\hat{V}(x) = 0$. This implies that for a specific
  energy $E$ in the continuum: $\psi(x) = \sin(\theta(x))/\sqrt{2 \pi k(x)}$,
  where $k(x)$ is the electron momentum which is proportional to $\sqrt{2E}$
  and it is related to $\theta(x)$ as $k(x) = d \theta (x) / dx$\relax
\mciteBstWouldAddEndPuncttrue
\mciteSetBstMidEndSepPunct{\mcitedefaultmidpunct}
{\mcitedefaultendpunct}{\mcitedefaultseppunct}\relax
\EndOfBibitem
\bibitem[mac()]{mac}
 The Mac\'ias procedure permits to normalise $L^{2}$-norm continuum states. In
  this method, for a specific energy $E$ in the continuum, we have: $\psi(x) =
  \psi_{L^{2}}(x) \sqrt{\rho(E) /2}$, where $\psi_{L^{2}}$ is the $L^{2}$-norm
  normalized state and $\rho(E)$ is the density of states evaluated numerically
  as the number of states per energy unit as $\rho(E_j) = 2 / (E_{j+1} -
  E_{j-1})$\relax
\mciteBstWouldAddEndPuncttrue
\mciteSetBstMidEndSepPunct{\mcitedefaultmidpunct}
{\mcitedefaultendpunct}{\mcitedefaultseppunct}\relax
\EndOfBibitem
\end{mcitethebibliography}

\end{document}